\documentclass[utf8]{frontiersinFPHY_FAMS} 

\setcitestyle{square} 
\usepackage{url,hyperref,lineno,microtype,subcaption}
\usepackage{siunitx}
\usepackage{caption}
\usepackage{subcaption}
\usepackage{float}
\usepackage{xcolor}

\usepackage{acronym}

\nolinenumbers

\definecolor{SVU}{rgb}{0.56, 0.0, 1.0}
\definecolor{BI}{rgb}{0, 0.5, 1.0}
\definecolor{MKo}{rgb}{0, 0.7, 0}

\def\keyFont{\fontsize{8}{11}\helveticabold }
\def\firstAuthorLast{Uchaikin {et~al.}} 
\def\Authors{Sergey V. Uchaikin\,$^{1,*}$, Jinmyeong Kim\,$^{2,1}$,  Çağlar Kutlu\,$^{1,2}$,  Boris I. Ivanov\,$^{1}$, Jinsu Kim\,$^{1}$, Arjan F. van Loo\,$^{3,4}$,  Yasunobu Nakamura$^{3,4}$, Saebyeok Ahn\,$^{1}$, Seonjeong Oh\,$^{1}$, Minsu Ko\,$^{2,1}$,  and Yannis K. Semertzidis\,$^{1,2,*}$}

\def\Address{$^{1}$Center for Axion and Precision Physics Research,  \ac{IBS}, Daejeon 34051, Republic of Korea \\
$^{2}$Department of Physics,  \ac{KAIST}, Daejeon 34141, Republic of Korea \\
$^{3}$\ac{RQC}, Wako, Saitama, 351–0198,Japan\\
$^{4}$Department of Applied Physics, Graduate School of Engineering, The University of Tokyo, Bunkyo-ku, Tokyo, 113-8656, Japan
}

\def\corrAuthor{Sergey V. Uchaikin, Yannis K. Semertzidis}
\def\corrEmail{uchaikin@kaist.ac.kr, yannis@kaist.ac.kr}

\begin{document}
\onecolumn
\firstpage{1}

\title[CAPP JPA ]{Josephson Parametric Amplifier based Quantum Noise Limited Amplifier Development for Axion Search Experiments in CAPP} 

\author[\firstAuthorLast ]{\Authors} 
\address{} 
\correspondance{} 

\extraAuth{}

\maketitle


\begin{abstract}

This paper provides a comprehensive overview of the development of flux-driven  \ac{JPAs} as Quantum Noise Limited Amplifier for axion search experiments conducted at the \ac{CAPP} of the Institute for Basic Science. It focuses on the characterization, and optimization of JPAs, which are crucial for achieving the highest sensitivity in axion particle detection. We discuss various characterization techniques, methods for improving bandwidth, and the attainment of ultra-low noise temperatures. JPAs have emerged as indispensable tools in CAPP's axion search endeavors, playing a significant role in advancing our understanding of fundamental physics and unraveling the mysteries of the universe.
\tiny
 \keyFont{ \section{Keywords:} axion dark matter, cavity haloscope, Josephson parametric amplifier, quantum noise limited amplifier, low temperature thermalization} 
\end{abstract}

\section{Introduction}
CAPP was established in October 2013, entering the realm of axion physics following the notable contributions of experiments like the  \ac{ADMX},  \ac{HAYSTAC},  etc.~\cite{article:ADMX-1, article:ADMX-2, article:HAYSTAC-1, article:HAYSTAC-2}. These experiments paved the way for axion research, and CAPP aimed to build on these achievements.  It focused on establishing itself as a significant player in the field by pushing the boundaries in all relevant parameters and expanding the detection sensitivity.

It is extremely difficult to observe axions because of their weak interactions with \ac{SM} fields. In the presence of a magnetic field, an axion with a mass of $m_a$ and momentum $p_a$ generates an electric field described by the Lagrangian,

\begin{equation}
    \mathcal{L}_{a\gamma}=-{g_{a\gamma\gamma}a_\text{DM}\textbf{B}_\text{ext}} \cdot\textbf{E},
\end{equation}
\noindent
 where $g_{a\gamma\gamma}$ is axion to two-photon coupling constant \cite{article:Cordona2016}; $a_\text{DM}$ is the axion dark matter field~\cite{article:Wilczek78, article:Weinberg78}${\textbf{B}_{ext}}$ and  ${\textbf{E}}$ are external magnetic and generated electrical field, respectively. Alternatively, it can be said that in the presence of a magnetic field field, the axion field mixes with the electromagnetic field of same frequency.
The resulting field has frequency $f_\text{ax}$ defined by the simple equation:

\begin{equation}
    f_{ax}=\frac{\sqrt{(m_ac^2)^2+(p_ac)^2}}{h}\approx\frac{m_ac^2}{h}.
\end{equation}
\noindent
Here, $h$ and $c$ represent the Planck constant and the speed of light in vacuum, respectively. The kinetic energy $(p_ac)^2$ of virialized axions is small, however, this kinetic energy broadens the line following a Maxwellian distribution with an equivalent quality factor of $Q_a\approx10^{6}$~\cite{Kim_2020}. To enhance the signal, we can adopt the haloscope concept introduced by Sikivie in 1983~\cite{article:Sikivie83} and first applied in 1987 by the authors of~\cite{article:DePanfilis1987}. Sikivie's proposed experimental setup involves a detector consisting of a microwave cavity resonator immersed in a strong magnetic field and cooled to cryogenic temperatures. The cavity plays a crucial role in resonantly accumulating the produced photons from the converted axion dark-matter field. When the frequency of the axion signal aligns with the resonance of the cavity, it may be observed using specialized low-noise readout equipment. 

\section{Haloscope}

The axion haloscope employs a cavity with a high-quality factor, positioned within a strong magnetic field for the detection of signals arising when the axion frequency aligns with the cavity's resonance. For the output signal power in haloscope experiments a larger magnetic field and cavity volume result in a greater conversion of axions into microwave photons (Eq.~\ref{eq:conv_power})~\cite{article:Brubaker2018}:

\begin{equation}
    P^{(a)}=\left[\left(\frac{g_{\gamma}\alpha f_a}{\pi}\right)^2\frac{\hbar^3c^3\rho_a}{m_a^2}\right]\left[\frac{\beta}{1+\beta}\omega_c\frac{1}{\mu_0}B_0^2VC_{010}Q_L\frac{1}{1+\left(\frac{2(f_c-f_\text{ax})}{\Delta f_c}\right)^2}\right],
    \label{eq:conv_power}
\end{equation}

\noindent
where $g_{\gamma}=(\pi f_a g_{\alpha\gamma\gamma})/\alpha$ is the coupling constant of the axion-photon interaction with values of $-0.97$ and $0.36$ for the \ac{KSVZ}~\cite{article:Kim1979, article:Shifman1980} and \ac{DFSZ}~\cite{article:Zhitnitsky1980, article:Dine1981} models, respectively; $\alpha$ is the fine structure constant; $f_a$, $\rho_a$ and $m_a$ are the decay constant, local halo density and mass of the axion dark matter, respectively.
$\beta$ is the coupling constant between the cavity and the readout system, with the term $\beta/(1+\beta)$ defining the fraction of the output power. $f_c=\omega_c/2\pi$ is the cavity resonance frequency;  $\mu_0$  is the magnetic permeability of free space; $B_0$ symbolizes the external magnetic field; $V$, $C_{010}$ and $Q_L$ are the cavity volume, form factor for a particular mode, in our case $\mathrm{TM}_{010}$, and the cavity loaded quality factor, {$\Delta f_c=f_c/Q_L$ }. The form factor characterizes the degree of overlap between the cavity TM$_{010}$ mode and the external magnetic field~\cite{article:Jeong_2022}:

\begin{equation}
    C_{010}=\frac{|\int\textbf{E}\cdot\textbf{B}_\text{ext}dV|^2}{\int\epsilon^{\prime}|\textbf{E}|^2dV\times\left\langle\textbf{B}^2_\text{ext}\right\rangle V}.
\end{equation}
\noindent
where $\epsilon=\epsilon^{\prime}+i\epsilon^{\prime\prime}$ is the relative permittivity within the cavity.
A higher quality factor ensures that converted microwave photons persist within the cavity for extended durations, thereby amplifying the signal. Reducing the physical temperature of the cavity and minimizing noise from the receiver chain, further enhance the efficiency of axion detection. 

In haloscope experiments, the typical cavity shape is cylindrical, fitting efficiently  into the design of a regular solenoid superconducting magnet, making optimal use of the magnet's bore volume. The cavity commonly utilizes the TM$_{010}$ mode, where the electric field aligns with the cylinder's axis and results in the maximum form factor. The central frequency of the cavity is primarily determined by its inner diameter. To precisely set the resonance frequency, one or more tuning rods, movable inside the cavity, are employed. Dielectric materials or highlyconductive metals are chosen for the tuning rod based on the desired frequency. 

To access the cavity signal, a movable antenna is employed to optimize the coupling $\beta$. As indicated in Eq.~\ref{eq:conv_power}, the output power strongly depends on cavity parameters $C_{010}$, $Q_L$, and $\beta$. This necessitates optimization of the cavity design, considering the impact of tuning rods on $C_{010}$ and aiming to avoid mode cross-coupling. The axion conversion power in the experiments performed at CAPP typically lays in a range between 10$^{-23}$ to 10$^{-22}$\,W. Such minuscule power levels impose stringent demands on the initial stage of our readout system.

The signal frequency is linked to the axion's mass $m_ac^2\approx hf_\text{ax}$, an attribute that is currently unknown. Therefore, it is necessary to scan across an extensive frequency spectrum to maximize the probability of capturing this  signal. The speed at which this frequency scan can be executed~\cite{Kim_2020}:

\begin{equation} \label{eq:1}
S\propto \frac{g_{\gamma}^4}{(SNR)^2}\eta\frac{1}{{T^2_\textrm{sys}}}{B_0}^4V^2{C_{010}}^2Q_L^2.
\end{equation}

\noindent
$SNR$ denotes the chosen signal-to-noise ratio (for the average of 5 and the threshold of 3.7 for a confidense level of 90\%); $\eta$ represents the  \ac{DAQ} efficiency; $T_\textrm{sys}$ is the system noise temperature. 

Consequently, the scanning frequency rate is inversely porportional to the square of the total system noise. Hence, one of our primary goals is to minimize it, thereby augmenting of our scanning speed. System noise emanates from multiple sources within our setup, encompassing noise generated by the cavity, other passive components and the amplifiers. To achieve the lowest noise levels, it is imperative to cool down all readout components to temperatures lower than, $T=(hf_\text{min})/k_B$, where $k_B$ are the Boltzmann constants, respectively, and $f_\text{min}$ denotes the minimum scanning frequency of the cavity.

The most advanced amplifiers are capable of approaching noise levels close to the quantum limit. This article is dedicated to the adaptation and application of such amplifiers.  In the following sections, we will elaborate on our measurement techniques for the JPAs, present their key parameters, and showcase both single JPA and multiple JPA readouts, including their respective noise characteristics. 

\section{Flux-driven JPA}

In the earlier stages of CAPP axion search experiments, such as CAPP-9\,T and CAPP-PACE  \cite{article:Jeong2020, article:Kwon2021},  we utilized state-of-the-art semiconductor amplifiers.
The InP \ac{HEMT} is renowned for its ability to provide the lowest noise temperature in cryogenic \ac{LNA} applications. We employed those manufactured by Low Noise Factory, specifically designed for cryogenic use \cite{LNF-LNC4_8F}. These devices exhibit exceptionally low noise temperatures, reaching about 1.2\,K when operated at temperatures below 10\,K for frequencies around 1\,GHz~\cite{Shleeh2012}. 
These experiments demonstrated CAPP's capability to create a high-quality \ac{RF} tunable cavity, maintain low temperatures, and operate a strong superconducting magnet in persistent mode~\cite{article:Jeong2020, article:Kwon2021}. It also provided invaluable experience in data collection and processing. However, readouts chain based on cryogenic HEMT amplifiers have noise levels several tens of times higher than the quantum noise limit. 
For this case it would take several decades to scan the range between 1 to 2\,GHz with DFSZ sensitivity.

\subsection{JPA design}

\begin{figure}[ht]
\begin{center}
\includegraphics[width=1\textwidth]{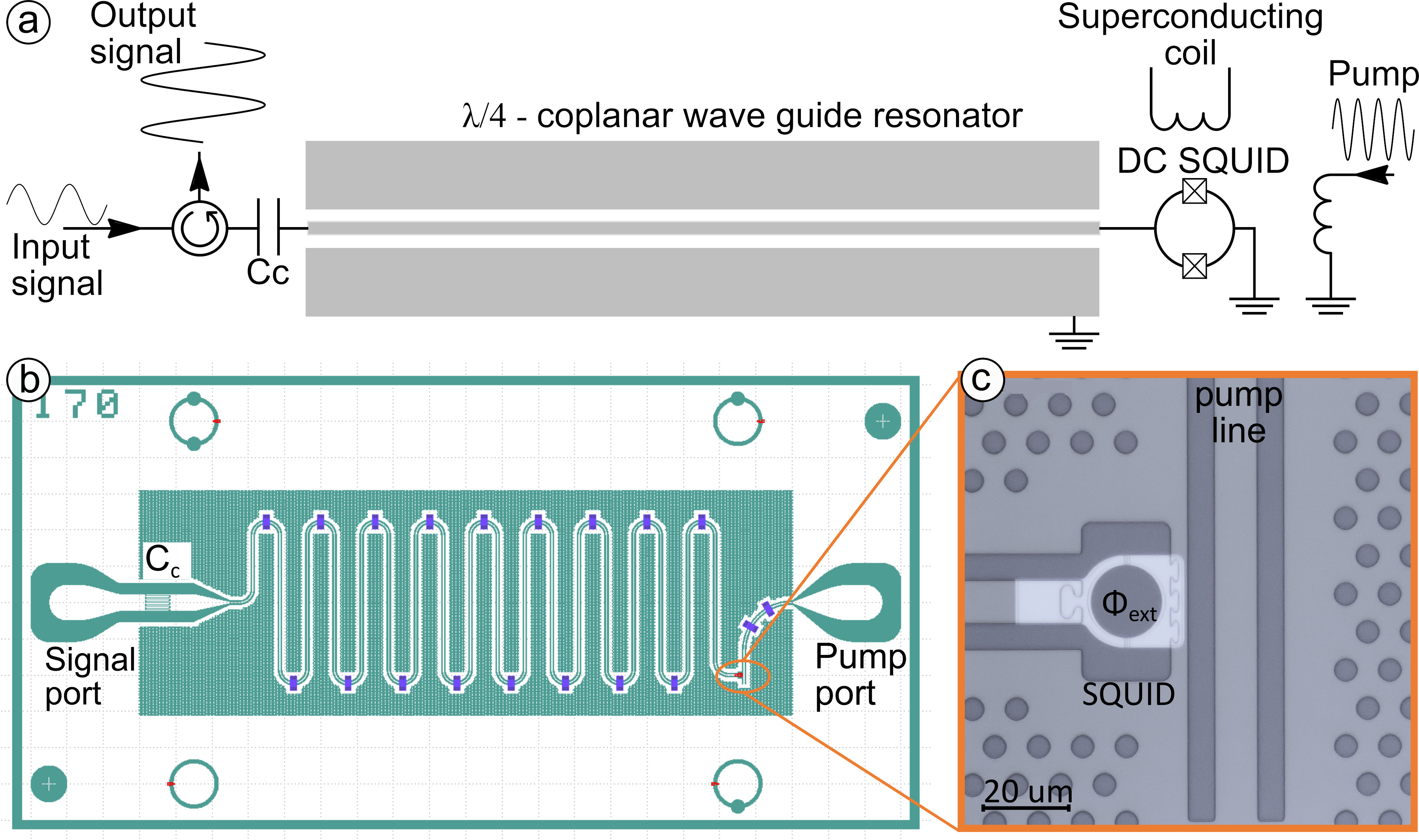}
\caption{\label{fig:FDJPA1} Flux-driven JPA. (a) Equivalent circuit diagram of the JPA. A flux bias defines the average inductance of the SQUID loop (the Josephson junctions are denoted as $\boxtimes$ in the loop), while the pump is modulating this flux at twice the frequency of the signal to be amplified. Since the signal's input and output utilize the same port, a cryogenic circulator is required to operate the amplifier. (b) JPA layout, where $C_c$ is the interdigitated capacitor. The $\lambda/4$ resonator is implemented as a meander line, whose length determining the maximum JPA frequency.  (c) Image of the SQUID (light structure in the center). The pump line is inductively coupled to the SQUID. }
\end{center}
\end{figure}

To achieve near-quantum-noise-limited amplification, we employ a flux-driven JPA designed and fabricated for these experiments in RIKEN and the University of Tokyo~\cite{article:Yamamoto08}. This JPA design consists of a $\lambda/4$ resonator terminated by a \ac{SQUID}, as depicted in Figure~\ref{fig:FDJPA1}. 
The main features of the devices are defined by using photolithography techniques on Nb film deposited on a 0.3-mm-thick silicon substrate. The SQUID, using Dolan-bridge type junctions~\cite{article:Dolan77}, is fabricated using electron-beam lithography followed by shadow evaporation of aluminium. Its inner loop has a diameter of roughly 16\,$\mu$m, and the junctions are designed to have identical critical currents of approximately 2\,$\mu$A. The SQUID connects the $\lambda/4$ resonator to ground. A pump line couples inductively to the SQUID, delivering a modulated flux at twice the resonance frequency. This pump signal is rejected by the $\lambda/4$ resonator as it only permits odd harmonics, and therefore does not couple into the signal input-output line. The pump line is shunted to ground close to the SQUID location, improving the inductive coupling. A static magnetic field generated by a superconducting coil is used to tune the frequency of the JPA.

Different JPAs were designed for amplifying different signal frequencies by changing the length of the $\lambda/4$ line. The instantaneous bandwidth of the device is designed by choosing the coupling capacitor size. A larger instantaneous bandwidth, when keeping the design otherwise constant, necessities in an increased pump strength for a certain gain and will have a reduced saturation power. The tunable bandwidth of these devices was limited by the small ratio of the SQUID inductance to the geometric inductance of the $\lambda/4$ \ac{CPW}, which has been addressed in more recent JPA designs to be used in future experiments. The devices were designed using COMSOL simulations to obtain the capacitance value of the interdigitated capacitor and the resonance frequency of the $\lambda/4$ CPW resonator. Numerical simulations using the negative resistance model~\cite{article:Sundqvist2014} were employed to predict the needed pump power, achievable gain, tunable range and instantaneous bandwidth.

The JPA is operated in the three-wave mixing mode~\cite{article:Roy2016}, where the frequencies of the pump $f_p$, signal $f_s$, and idler $f_i$ are related by:

\begin{equation} \label{eq:f_p}
f_p = f_s + f_i.    
\end{equation}

The JPA chips have dimensions of 2.5$\times$5$\times$0.3\,mm$^3$, see Fig.\,\ref{fig:FDJPA1}(b), and are fixed on top of a copper block. A \ac{PCB} to connect microwave lines to the input and pump lines is fixed on the same copper block and wirebonded to the chip. This copper block is encased within a copper base for the DC bias coil and thermally bonded to it. When not in use, this assembly is stored in an antistatic plastic box at room temperature within an evacuated desiccator. The detailed sample holder design is described elsewhere~\cite{article:PRX2024}.

\section{JPA measurement setup}

\subsection{RF setup}

\begin{figure}[ht]
\begin{center}
\includegraphics[width=1\textwidth]{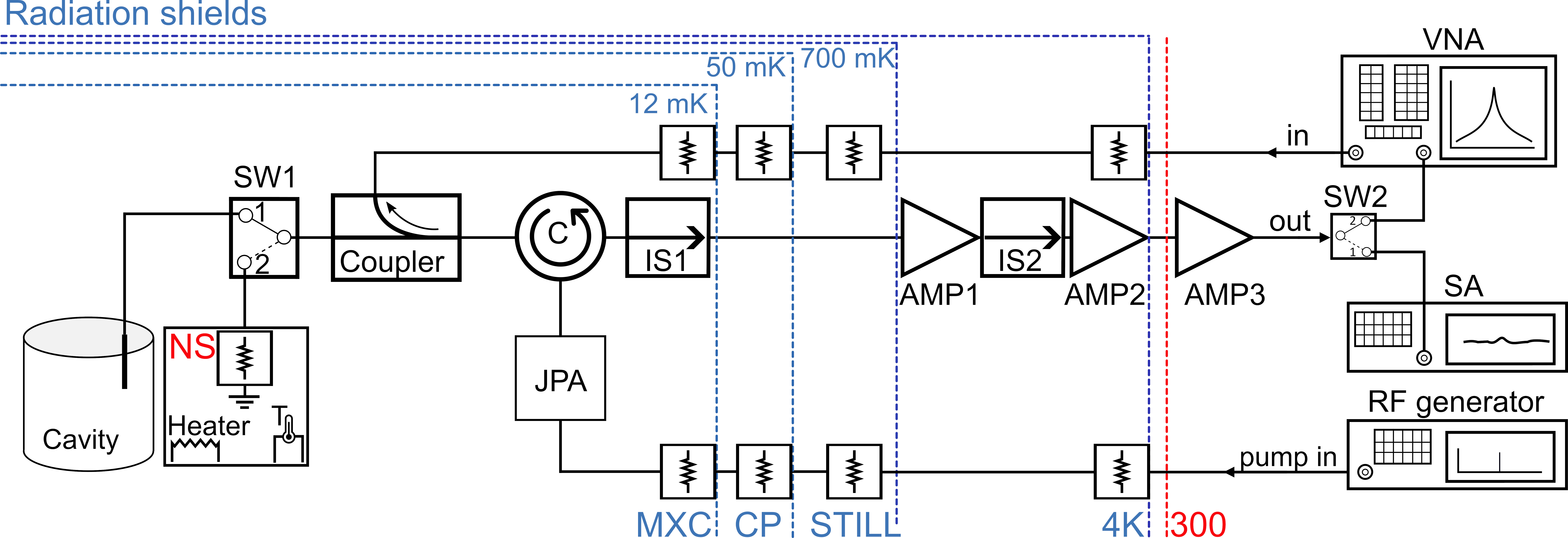}
\caption{\label{fig:Fig2_main_diagram} JPA readout diagram with either a cavity or a noise source at the input. Components arranged from left to right: noise source for tests or
cavity for experiment run, SW1: RF switch to toggle between JPA-test and axion-experiment modes; the Coupler: ``cold'' directional coupler for injecting test signals into the JPA; components C and IS1 are a cryogenic RF-circulator and an RF isolator; AMP1 and AMP2: cryogenic HEMT amplifiers; IS2: RF isolator between HEMTs; AMP3: room temperature amplifier; SW2: RF switch between \ac{VNA} and \ac{SA};  VNA, SA, and RF generator.
}
\end{center}
\end{figure}

In Figure~\ref{fig:Fig2_main_diagram}, one can observe the basic setup for testing and for the axion search experiment, featuring various temperature stages, including 300\,K, 4\,K, still, \ac{CP}, and \ac{MXC}  of the \ac{DR}. This design is adaptable to different types of DRs and the nature of the experiment. It offers inputs for measuring various parameters such as tuning frequency range, gain, instantaneous bandwidth, and noise temperature of the JPAs. All RF connections from the 300\,K flange to the 4K plate are made with CuNi coaxial 0.047-inch-diameter cables, provided either by \ac{BF}~\cite{Bluefors} or made in-house. Below 4K, some of the RF lines are done with the same cable, while the signal lines are constructed using superconducting NbTi 0.047-inch-diameter cables to minimize heat load and reduce losses. Each input line is equipped with a series of attenuators, thermally anchored to each temperature stage of the fridge, to reduce Johnson noise originating from higher temperature stages. 
The attenuator values are chosen by considering the equation below:
\begin{equation} \label{eq:T_out}
 T^{\text{out}}_\text{att} = T^{n-1}_\text{att}A + T^n_\text{att}(1-A),
\end{equation}
\noindent
where \( T^{\text{out}}_\text{att} \) is the output noise temperature of the attenuator, \( T^{n-1}_\text{att} \) is the output noise temperature of the attenuator on the previous temperature stage, \( T^n_\text{att} \) is the physical temperature of the attenuator, and \( A \) is the attenuation coefficient.
Here, we assumed that the physical temperature of the attenuator equals the temperature of the corresponding stage of the fridge. This assumption is appropriate since the dissipation and self-heating are negligible when small power signals are involved.
The  {RF} chain was installed on the MXC plate and consists of a range of cryogenic microwave components, including couplers, circulators, and isolators.
Prior to their use at\,m\si{\kelvin} temperatures, each component was preliminary characterized on a cold testing bench. To control the bias current $i_b$ we use a Yokogawa GS200 current source~\cite{Yokogawa}

\subsection{DC wiring, cold DC filters and switch boxes}
To test our JPAs and other RF and DC components, we set up a dedicated test system using a BF LD400 dry fridge~\cite{Bluefors-LD400}. To enhance our testing capabilities, we implemented a set of 32 differential lines, each individually shielded, into the fridge. These lines are constructed using twisted pairs, each placed within a Teflon matrix and equipped with its own CuNi braided shield. For pairs connecting the \ac{RT} flange to the 4\,K stage, we utilized cables with 0.1 mm diameter brass wires. For connections between the 4\,K stage and the MXC, twisted pairs were made from copper nickel coated 0.1\,mm NbTi wires. The cables were thermally connected to each temperature stage through their CuNi shields. On the 4\,K stage, they were thermally connected via a specialized interconnection box, providing a transition between brass and superconducting wires. At a base temperature of approximately 11\,mK on the MXC, the estimated maximum heat load from wires connected to the upper stage does not exceed 0.8 $\mu$W.

\begin{figure}[ht]
\begin{center}
\includegraphics[width=.9\textwidth]{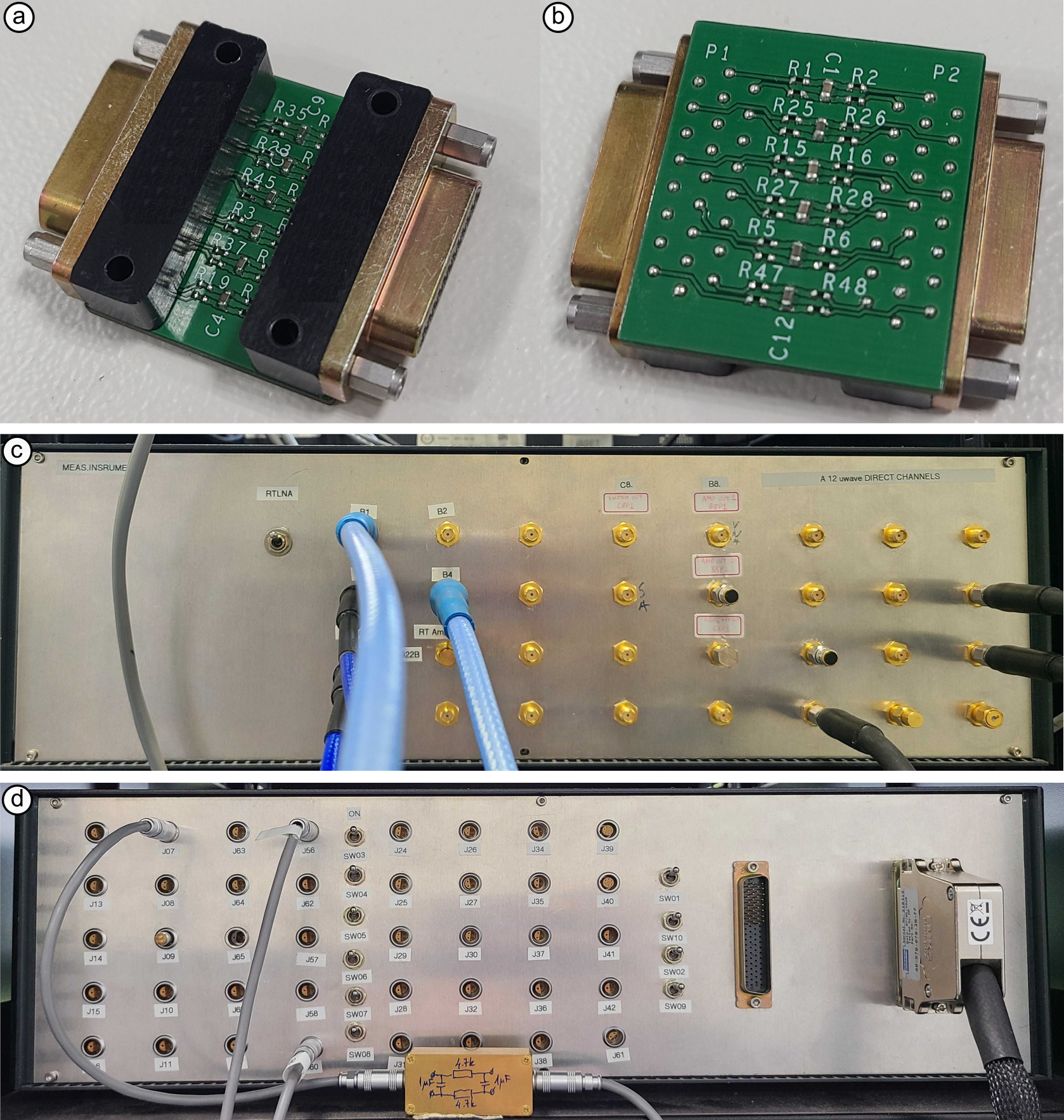}
\caption{\label{fig:RC-filters} DC and RF wiring. (a) Top and (b) bottom views of the cryogenic filter block for the BF fridge. The block contains 12 differential T-type RC filters designed for standard BF DC lines and is mounted on the 4\,K stage of the fridge. (c) DC and (d) RF boxes of the test fridge BF4 located at room temperature.}
\end{center}
\end{figure}

In the fridge DC lines, we incorporated DC filter assemblies at the 4\,K stage. Each of our filter assemblies contained  12 differential T-type RC  filters. In the lines used to provide power for the HEMT amplifiers, the RC filters are replaced with LC filters in order to reduce resistive losses. These filters are constructed on a two-layer PCB, utilizing both sides as ground planes to minimize interference and enhance thermalization, see [Fig.~\ref{fig:RC-filters}(a)].

To safeguard the connectors on the top of the BF fridge from potential damage due to frequent connection and disconnection during testing, we have designed two specialized boxes for DC and RF connections~[Figs.~\ref{fig:RC-filters}(c) and~(d)]. These boxes are permanently connected to the fridge, and all connections for setup, testing, and experiments are made within these boxes. The RF box also houses an RF amplifier, while the DC box features switches controlled by a 40-411-001 PXI 64 Channel Relay Driver Module~\cite{Pickering} based on a PXES-2590 chassis~\cite{ADLINK}, enabling flexible configurations for testing purposes.

\subsection{Noise source}\label{sec:NoiseSource}

\begin{figure}[ht]
\begin{center}
\includegraphics[width=.98\textwidth]{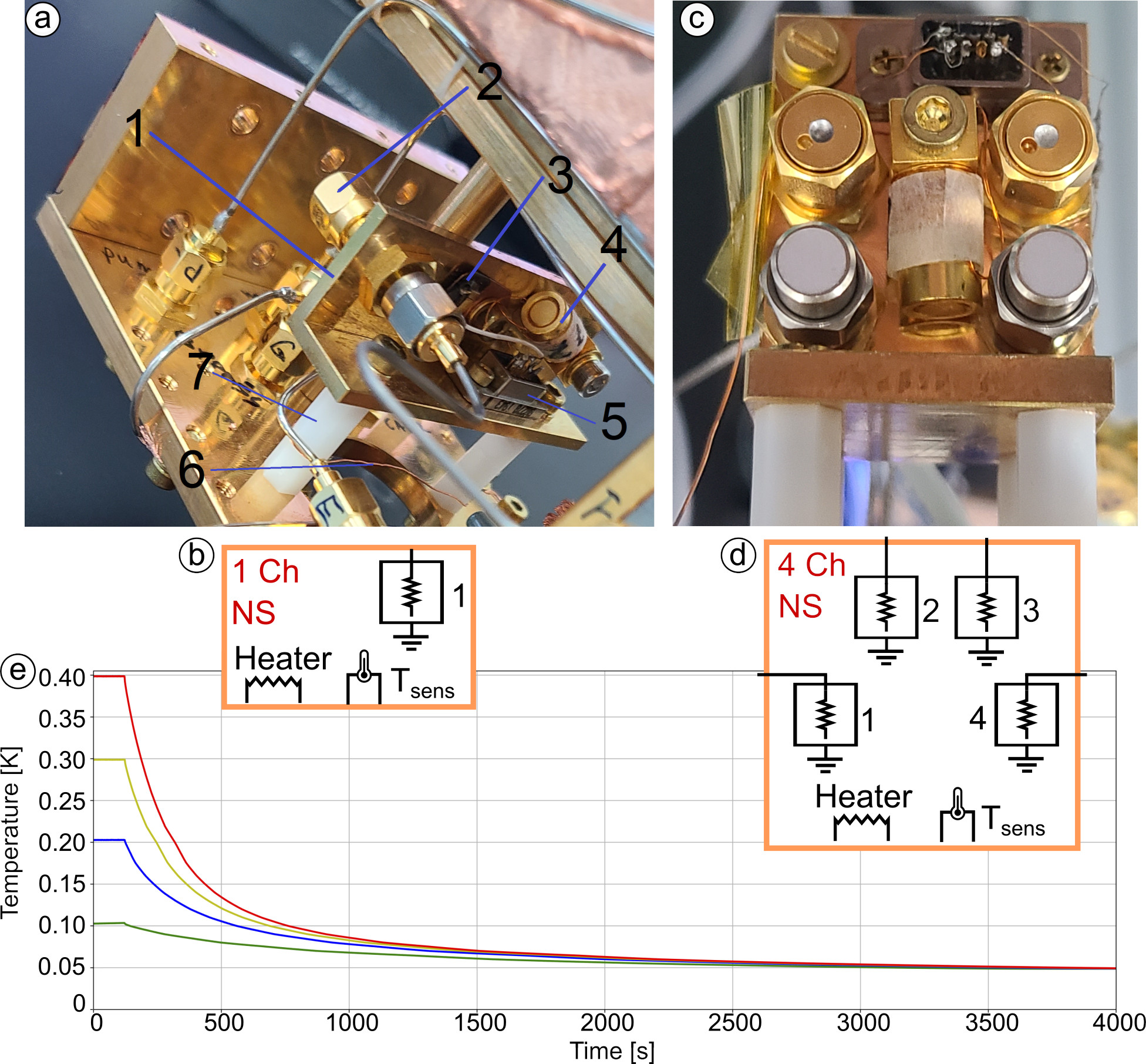}
\caption{\label{fig:NoiseSource} Cryogenic NS. (a) One-channel NS components on the MXC assembly: 1 - enclosure, 2 - 50\,$\Omega$ terminator, 3 - heater, 4 - thermometer, 5 - connector, 6 - thermalization wire, 7 - isolating plastic stand. (b)~Equivalent circuit of the one channel NS. (c) Four-channel NS and (d) its equivalent circuit~\cite{article:Ivanov23-LT29}. (e) Cooling of a single-channel NS heated up to 100, 200, 300 and 400\,mK. The NS is thermally connected to the MXC with the temperature stabilized at 50\,mK.}
\end{center}
\end{figure}

Our setup  incorporates a \ac{NS}, facilitating the measurement of noise parameters for the JPAs and cryogenic HEMT amplifiers.
The NS is a separate assembly comprising a wideband matched 50\,$\Omega$ terminator, fixed on a bulkhead, labelled as ``2" in Fig.~\ref{fig:NoiseSource}(a), a heater ``3'', and a thermometer ``4''. These components have been integrated onto a copper fixture (1). To enhance thermal connections, the enclosure was plated with gold of 6\,$\mu$m thick.
The NS equivalent circuit is shown in Fig.~\ref{fig:NoiseSource}(b).

This NS has been designed for placement on the MXC plate of dry BF or wet Leiden dilution refrigerators. Our design prioritizes compact dimensions, minimizing the footprint on the MXC plate while ensuring adequate heat capacity for thermal stabilization. Bulkhead SMA adapters were employed to establish connections between the 50\,$\Omega$ cryogenic wideband matched loads (2) and the measurement chain. These adapters, in conjunction with the wideband XMA cryogenic 50\,$\Omega$ terminations boasting a maximum \ac{VSWR}
of 1.15, guarantee 50\,$\Omega$ matched connections across the frequency range up to 18 GHz.
 At temperature $T$, the source generates Johnson-Nyquist noise with the noise temperature $T_n$ in accordance with the Nyquist theorem:

\begin{equation}
    T_n=\frac{h f}{k_B}\left(\frac{1}{2}+\frac{1}{e^{h f/k_BT}-1}\right) \label{eq:T_n}.
\end{equation}

For the interconnections between the NS and the rest of the measurement chain, we utilized 0.047-inch-diameter NbTi microwave coaxial cables. Three superconducting NbTi twisted-pair cables within braided CuNi shields were employed to connect the temperature sensor (4) and heater (3) to the micro-Sub-D-9 connector (5) of the NS. The four-point temperature measurements were conducted using a calibrated ruthenium-oxide temperature sensor. The heater was crafted using a 100\,$\Omega$ precision resistor, with its housing carefully removed. It was then securely attached to the NS assembly using epoxy glue. The Lake Shore Cryotronics 372 AC resistance bridge and temperature controller enabled precise temperature measurements and control.
NbTi superconducting twisted pairs and RF cables were utilized to minimize losses and thermal conductance between the MXC and the NS connectors. Additionally, the NS assembly was securely attached to the MXC plate using two 20-mm-long plastic spacers (7). Since both DC and RF connections to the NS were made using superconducting cables, thermal anchoring to the MXC plate primarily relied on a weak link created with copper wire (6). 

Through a \ac{PID} controller, the NS temperature could be adjusted from 50\,mK to 1\,K without affecting the MCX plate temperature.
Initially, we employed a single-channel NS in our setup. However, as our testing requirements evolved and the demand for increased testing speed became apparent, we  transitioned to a four-channel NS configuration~\cite{article:Ivanov23-LT29}, all housed within a single enclosure as shown in Fig.~\ref{fig:NoiseSource}(c).
The four channel NS equivalent circuit is shown in Fig.~\ref{fig:NoiseSource}(d).
This allowed us to significantly accelerate our testing processes. 

The time it takes for the temperature of the NS to stabilize depends on the heat capacitance of the NS itself, along with all the components mounted on it, and its thermal coupling to the MXC. This thermal  conductance of the coupling is determined by the length and diameter of the copper magnet wire, the annealing of the wire, and the quality of the contacts between the wire, the NS enclosure on one side, and the MXC plate on the other.
To ensure efficient thermal transfer, we took careful steps to establish good contacts. We removed the insulation clad from the wire and pressed the wire against the enclosure and plate using brass screws. We attempted to enhance contact reliability by introducing clamp contacts on both sides of the wires, but this approach led to a significant reduction in thermal conductivity, preventing us from lowering the temperature of our NS below about 180 mK. We attribute this to the introduction of substantial mechanical defects in the crimped connections, resulting in a significant increase in thermal resistance. A similar effect has been observed in other experiments~\cite{Priv_Comm_Rybka}. After returning to our initial connection method, the thermal connection was restored.

In Fig.~\ref{fig:NoiseSource}(e), the relaxation of the NS temperature after it was set to target temperatures is shown. The relaxation curve  deviates from an exponential shape due to the strong non-linearity of thermal conductance and heat capacities at millikelvin temperatures.

\subsection{Shielding of JPAs and RF components from  magnetic fields}

According to Eq.~\ref{eq:1}, the scanning speed of axion haloscope searches is strongly determined by the magnet's strength and size, making it one of the pivotal factors in these experiments. The presence of such a magnet also generates a large magnetic field in its vicinity. Contemporary magnets incorporate compensation coils to establish a reduced magnetic field zone near the magnet. Nevertheless, the available space for this purpose is constrained by the size of the fridge. Consequently, it may not be sufficient to accommodate all sensitive components. Our superconducting magnets, which generate a magnetic field ranging from 8 to 18\,T, with an aperture up to 320\,mm, produces in a magnetic field strength of below 20\,mT around the RF-readout components. As such, magnetic shielding is necessary.

\subsubsection{RF switch shielding}

\begin{figure}[ht]
\begin{center}
\includegraphics[width=1\textwidth]{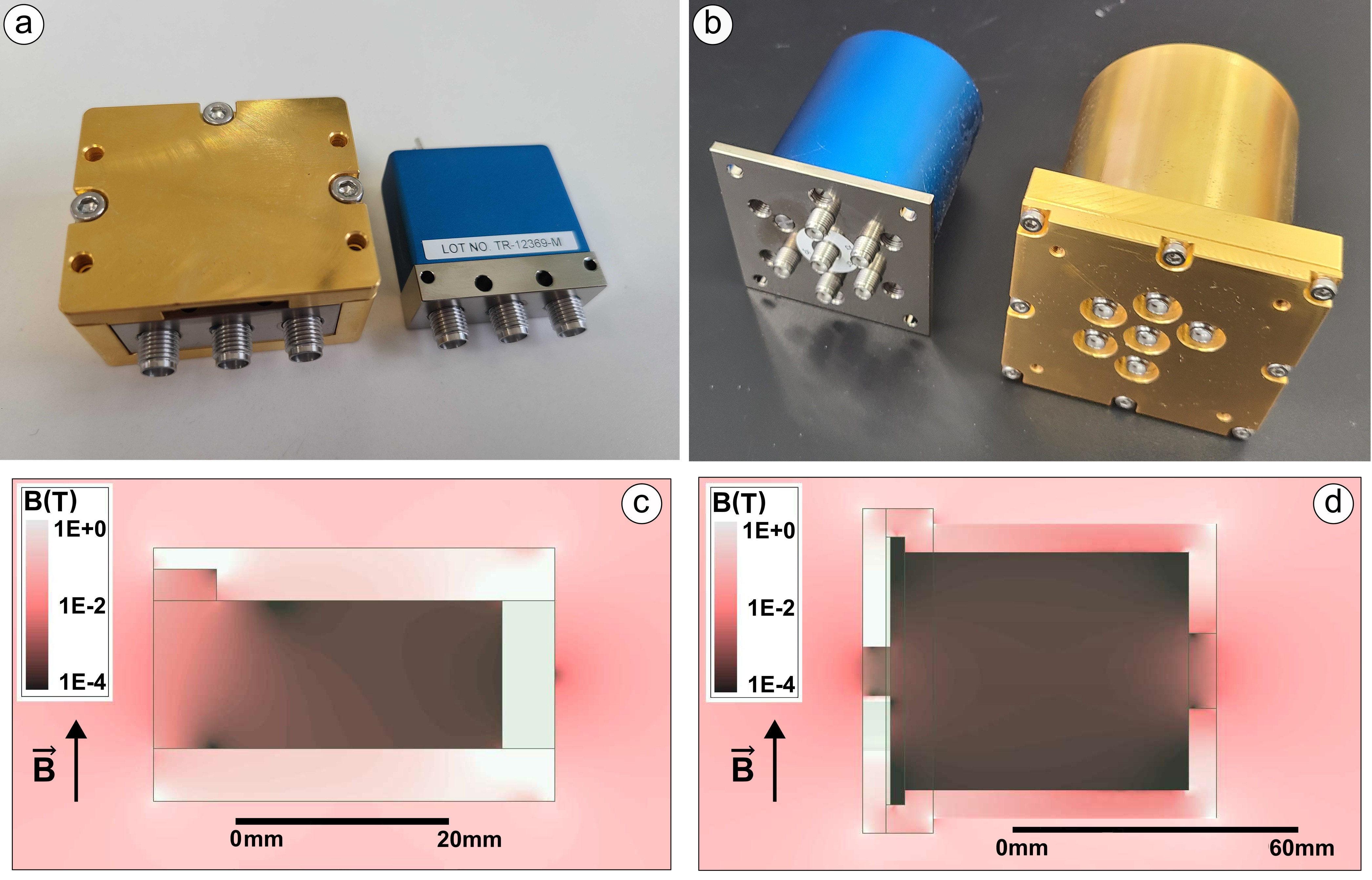}
\caption{\label{fig:switch_shield} Two types of RF switches used at CAPP with and without magnetic shields. (a) Three-position RF switch. (b) Five-position RF switch. (c) and (d) ANSYS~\cite{ANSYS} simulations of magnetic field strength under an external vertical magnetic field of 100\,mT for the three-position and five-position RF switches, respectively.}
\end{center}
\end{figure}

Within our experimental setup, we face the challenge of potential magnetic field interference affecting a range of critical components, including circulators, isolators, HEMT amplifiers, and notably, the JPA. To surmount this challenge, we have used shielded components, like circulators and isolators.
For the unshielded components that can only function in low magnetic field environments, like RF switches~[Figs.~\ref{fig:switch_shield}(a) and (b)], special shields were implemented. 

\subsubsection{Onion shield}\label{sec:OnionShield}

\begin{figure}[ht]
\begin{center}
\includegraphics[width=0.98\textwidth]{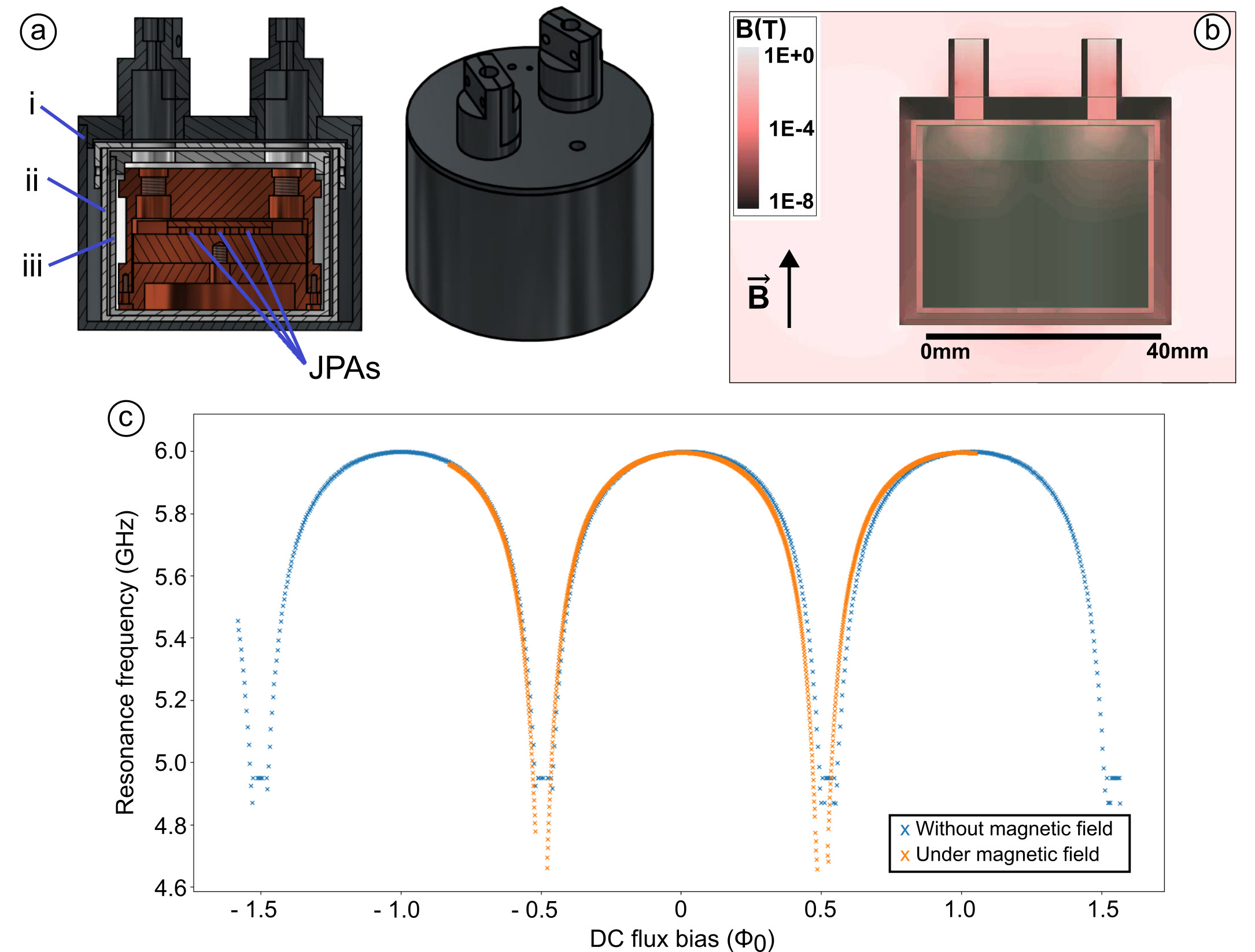}
\caption{\label{fig:Onion_shield} Onion shield. (a) Three-layer design of the shield.~(i)~Niobium or NbTi superconducting shield, (ii)~Cryophy shield, (iii)~inner superconducting Al shield. The JPA locations are indicated by the arrows. (b) ANSYS~\cite{ANSYS} simulation of the magnetic field including only the two outer layers of the \ac{OS} under an external field of 50\,mT and temperature of \text{100\,mK}. (c) Resonance frequency $f_r$ of the JPA in the passive mode depending on the applied DC flux bias, both in the presence and absence of an applyed vertical residual magnetic field of the superconducting magnet. The field strength at the components location is expected to be lower that 50\,mT. The measured shielding factor corresponding to these conditions ranges between $10^5$ and $10^6$.} 
\end{center}
\end{figure}

JPAs contain the most sensitive component to magnetic fields, i.e., the SQUID. In order to safeguard them against the influence of magnetic fields, we have developed a three-layer shielding system~\cite{article:Uchaikin23-LT29}, referred to as the \ac{OS} due to its nested design, see Fig.~\ref{fig:Onion_shield}(a). 
The performance of the OS results from its design and the properties of the materials used.

The shield has three layers, each with its own function to block and control magnetic fields. The outermost layer [(i) in Fig.~\ref{fig:Onion_shield}(a)], constructed with Nb or Nb alloy superconducting materials, has a high critical magnetic field value, and the external magnetic field is shielded by it only at cryogenic temperatures. The second layer (ii), consisting of a ferromagnetic material with high magnetic permeability~\cite{cryophy,ARPAIA2019514}, shields magnetic fields penetrating the outer shield through the structural openings. Finally, the innermost layer (3) is also made of superconducting aluminum with a critical temperature of about 1.2\,K.

\begin{table}[h]
  \caption{Onion shield layer properties.}
    \centering
    \begin{tabular}{|c|c|c|c|}
    \hline
         Layer  & Outer & Middle & Inner\\
         \hline
        Material & Nb or NbTi & CRYOPHY & Al \\
    \hline
        Superconducting transition, K & 9.2-11 & N/A & 1.2 \\
    \hline 
    Maximum magnetic field, T & \textgreater 0.1 & $\approx10^{-4}$ (saturation) & \textless 0.01\\
    \hline
    \end{tabular}
      \label{tab:Onion_shield}
\end{table}

At non-cryogenic temperatures, the OS acts like a regular magnetic shield due to its second layer made of ferromagnetic material. In fields with a magnitude close to the Earth's field, the calculated field inside the OS is a few tens of nT. During the initial cooling, the outermost layer undergoes a superconducting transition at the critical temperature (for our material choice it is between 9.2 and 11\,K). During the transition from the normal to the superconducting state, the outer layer of the OS “freezes” the Earth’s magnetic field inside the shield. Throughout this process, the magnetic field inside the second shield, composed of a ferromagnetic material, remains predominantly unaffected.

As the cooling process continues and the temperature approaches the critical point for the innermost 
layer~(iii) of the shield, its phase transition occurs. This innermost layer of the shield, made of superconducting aluminum, freezes the residual magnetic field that penetrated through the previous layers, i.e. a few tens of nT.

Since the first layer of the OS is in a superconducting state, while the field strength of the magnet increases, a persistent current appears on the surface of the first layer of the OS due to Meissner effect. This current freezes the magnetic field within this layer. However, some magnetic field penetration is inevitable through the openings required for cable routing and thermalization.

The second shielding layer plays a pivotal role in substantially reducing the penetration of the magnetic field into the first layer of the OS. However, again due to cable holes, there is a small penetration through it. This change in the residual field of the ferromagnetic shield is further shielded by the aluminum layer. Persistent currents induced on the surface of the third shield counteract minor field fluctuations within the second layer, ensuring that these fluctuations do not impact the magnetic field within the OS internal volume. Consequently, this configuration effectively shields the internal volume from the influence of external magnet field variations. The magnetic field within the third layer of the shield undergoes only marginal changes, even when the magnet is on, remaining within a range of 100\,nT. This is confirmed by both simulation and experiments. Simulations in Fig.~\ref{fig:Onion_shield}(b) and measurement, in Fig.~\ref{fig:Onion_shield}(c), showed consistent results.
The measurements were done with a 5.8\,GHz JPA.

\section{JPA characterization}

For all of our devices, we conduct thorough testing procedures. Since the JPA functions as a resonant amplifier, our objectives are to determine the tunable range, explore the relationships between gain, instantaneous bandwidth, and noise as a function of pump power and detuning frequency, and establish the input signal range within which the amplifier does notsaturate. The key steps in our characterization process include:

\begin{enumerate}
    \item Passive resonance identification by measuring the resonance frequency as a function of the DC flux.
    \item Gain parametric map (paramap) construction (gain $G$ vs. pump power $P$ and detuning $\delta = f_p/2 - f_r$ per bias current $i_b$), illustrating the gain concerning both pump power and detuning frequency.
     \item The 1 dB gain compression point ($P_{1\text{dB}}$) measurement in order to determine the input saturation power of the amplifier.
    \item Noise paramap generation by measuring the noise temperature $T_n$ vs. pump power and detuning.

\end{enumerate}


\subsection{Resonance frequency vs DC flux bias current}

In the passive regime, where no pump signal is applied, the JPA  operates as a weakly coupled quarter-wave resonator, shorted to  ground through a non-hysteretic SQUID, and coupled to the outside circuit via a capacitor $C_c$ [see Fig.~\ref{fig:FDJPA1}(a)].
When subjected to a small input signal, the SQUID remains in a superconducting state, effectively acting as a flux-dependent inductor. Consequently, we can treat the JPA as a quarter-wavelength resonator, with its effective inductance tunable by adjusting the flux $\Phi_\text{ext}$ going through the SQUID loop,

\begin{equation} \label{eq:L_eff}
L_\text{eff} = \frac{\Phi_0}{2\pi I_c|\cos{(\pi\Phi_\text{ext}/\Phi_0)|}},    
\end{equation}

\noindent
where $\Phi_0=2.067\cdot10^{-15}$\,Wb represents the flux quantum, and $I_c\approx I_{c1} \approx I_{c2}$ signifies the critical currents of junction~1 and junction~2, respectively. 

The operating frequency of the JPA is determined by both the physical dimensions of the transmission line and the inductance of the SQUID. To adjust this frequency within a specific range, we utilize an inductively coupled DC flux bias coil. This coil is constructed as a single-layer coil wound with 0.05\,mm copper-coated NbTi wire on a copper core, featuring an inner diameter of 32.5\,mm and a length of 11.5\,mm. It comprises 200 windings and has an approximate resistance of 200~$\Omega$s at room temperature.

\begin{figure}[ht]
\begin{center}
\includegraphics[width=1\textwidth]{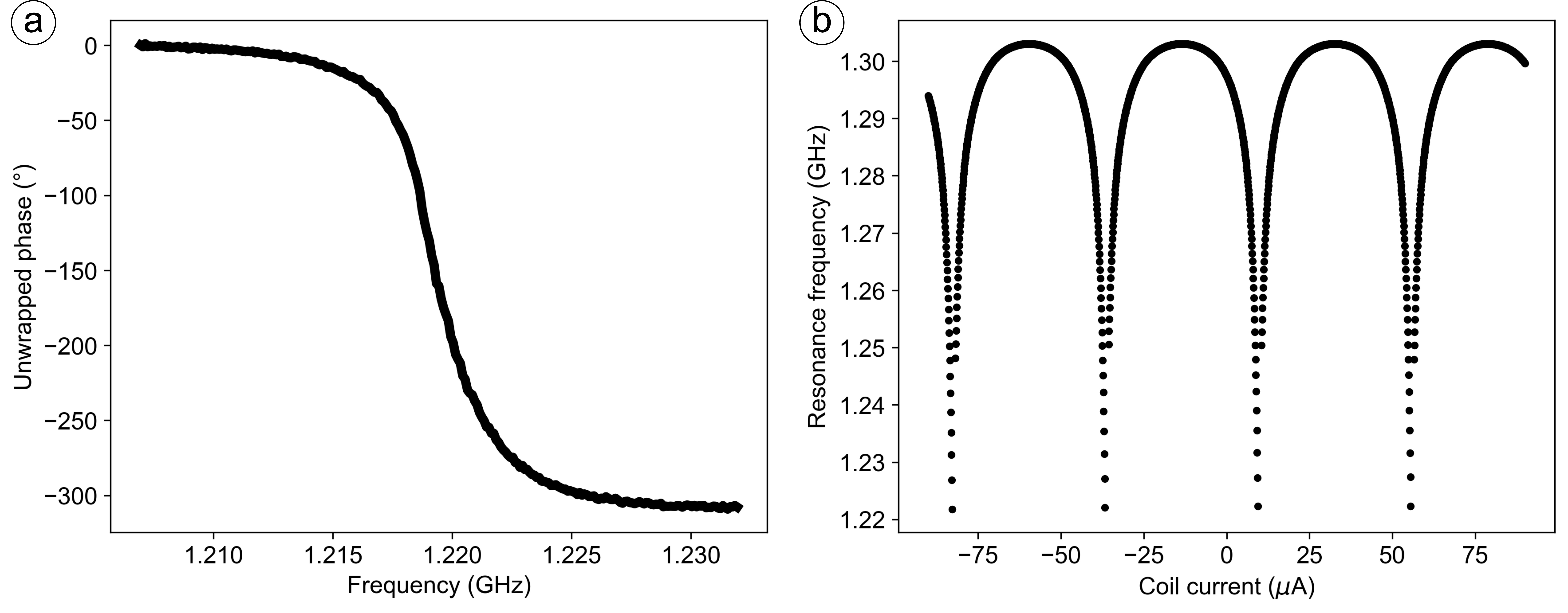}
\caption{\label{fig:Passive_JPA} The passive resonance of the JPA. (a) Dependence of $S_{21}$ phase vs. frequency for a JPA measurement in the circuit in Fig.~\ref{fig:Fig2_main_diagram} for a fixed current when the pump signal is switched off. (b) Relationship between the resonance frequency and the DC flux bias for a JPA. }
\end{center}
\end{figure}

The impedance of a quarter-wavelength resonator becomes finite and real at resonance, ideally zero. As a result, the reflection coefficient exhibits a dip at resonance. However, in the low loaded quality factor regime, it results in a small and broad dip spread over a wide frequency range (approximately 100--200 MHz). When measuring the JPA through various components, each with its own characteristics, distinguishing this dip becomes challenging. On the other hand, phase measurement is comparatively easier because the resonance induces a 360-degree phase shift, whereas the phase shifts of other components do not exhibit such a strong frequency dependence.
For this reason, we utilize phase-frequency measurements. We examine the JPA by sending a tiny signal towards it and recording the reflected signal. To separate the incoming and outgoing waves, a circulator is employed. The signal from the JPA is then amplified using two low-noise cryogenic HEMT amplifiers, AMP1 and AMP2, and RT amplifier AMP3, see Fig.~\ref{fig:Fig2_main_diagram}. The output signal recorded at the VNA input is influenced by several frequency-dependent components within the chain, resulting in the observed signal profile depicted in Fig.~\ref{fig:Passive_JPA}(a).
The reflection signal of the JPA is obtained through $S_{21}$ measurements using a VNA in the configuration shown in Fig.~\ref{fig:Fig2_main_diagram}. Subsequently, the phase information of this spectrum is employed to extract the resonance frequency, utilizing a parameter fit based on Ref.~\cite{Krantz_2013}. The resonance frequency is measured as a function of the coil current, as shown in Fig.~\ref{fig:Passive_JPA}(b). The measurement reveals that the minimum observable resonance frequency was below 1.25\,GHz, while the maximum reached 1.33\,GHz. However, at lower frequencies, the JPA exhibits large sensitivity to flux noise, attributed to the higher $\frac{\partial f_r}{\partial i_b}$, and therefore this device was only used above 1.26\,GHz.

The phase of reflected signal is given by:

\begin{equation}
    \text{arg}(S_{11})=\text{arctan}\left(\frac{2\Gamma_0\delta f}{\delta f^2+\left[\Gamma^2_R-\Gamma^2_0\right]}\right)
    \label{eq:phase}
\end{equation}

\noindent
Here,  $\delta f = f - f_r$ indicates the frequency deviation from the resonance frequency $f_r$. $\Gamma_R = {f_r}/{(2Q_i)}$ signifies the half-width of the resonance curve, determined by the internal quality factor $Q_i$. It signifies the bandwidth of the resonator attributed to its internal losses. Similarly, $\Gamma_0 = {f_r}/{(2Q_e)}$ denotes the half-width of the resonance curve, determined by the external quality factor $Q_e$.  It denotes the bandwidth of the resonator, accounting for both intrinsic and coupling losses.

\subsection{JPA gain measurement}\label{sec:Gain}

Upon applying a pump signal with frequency \( f_p \approx 2f_s \), the three-wave mixing mode obeys  Eq.~\ref{eq:f_p}. Using the JPA in three-wave mixing mode causes an amplified signal and an idler mode to appear symmetrically around \( {f_p}/{2} \). In the circuit shown in Fig.~\ref{fig:Fig2_main_diagram}, we estimate the JPA gain through the change in amplification in the output by comparing transmission measurements with the JPA in both on and off states. 
This establishes the baseline measurement, denoted as \( S_{\text{off}} \). A separate measurement is performed with a microwave short circuit connector replacing the JPA to assess the accuracy of this baseline. The comparison indicates that the baseline obtained from this off-resonance measurement is at most  \(0.2\,\text{dB} \) less efficient than if an ideal reflection was achieved.

Next, the JPA is tuned to the desired working point, and a subsequent transmission measurement \( S_{\text{on}} \) is taken. From these two measurements, the power gain is determined via equation:

\begin{equation}
   G_J = 10 \log_{10} \left| \frac{S_{\text{on}}}{S_{\text{off}}} \right|^2.  \label{eq:G_J} 
\end{equation}

\noindent
To find the optimal working point, we measure a ``paramap", which represents the dependence of \( G_J \) measured at \( f_s = {f_p}/{2} + 1\, \text{kHz} \) on the pump frequency \( f_p \) and pump power \( P_p \). The \( 1\, \text{kHz} \) offset is enough to avoid the degenerate mode of the JPA. This measurement is repeated for all tuning frequencies of the JPA by varying \( i_b \).

To determine the operational points across the entire frequency range, we devised a two-step procedure. Optimization included finding values of \( f_p \) and \( P_p \) for minimum noise and acceptable gain  across the entire operational frequency range controlled by the \( i_b \) value. Our desired gain falls within the range of 15 to 20 dB, which minimizes the noise contribution of the subsequent HEMT amplifiers.

\begin{figure}[ht]
\begin{center}
\includegraphics[width=1\textwidth]{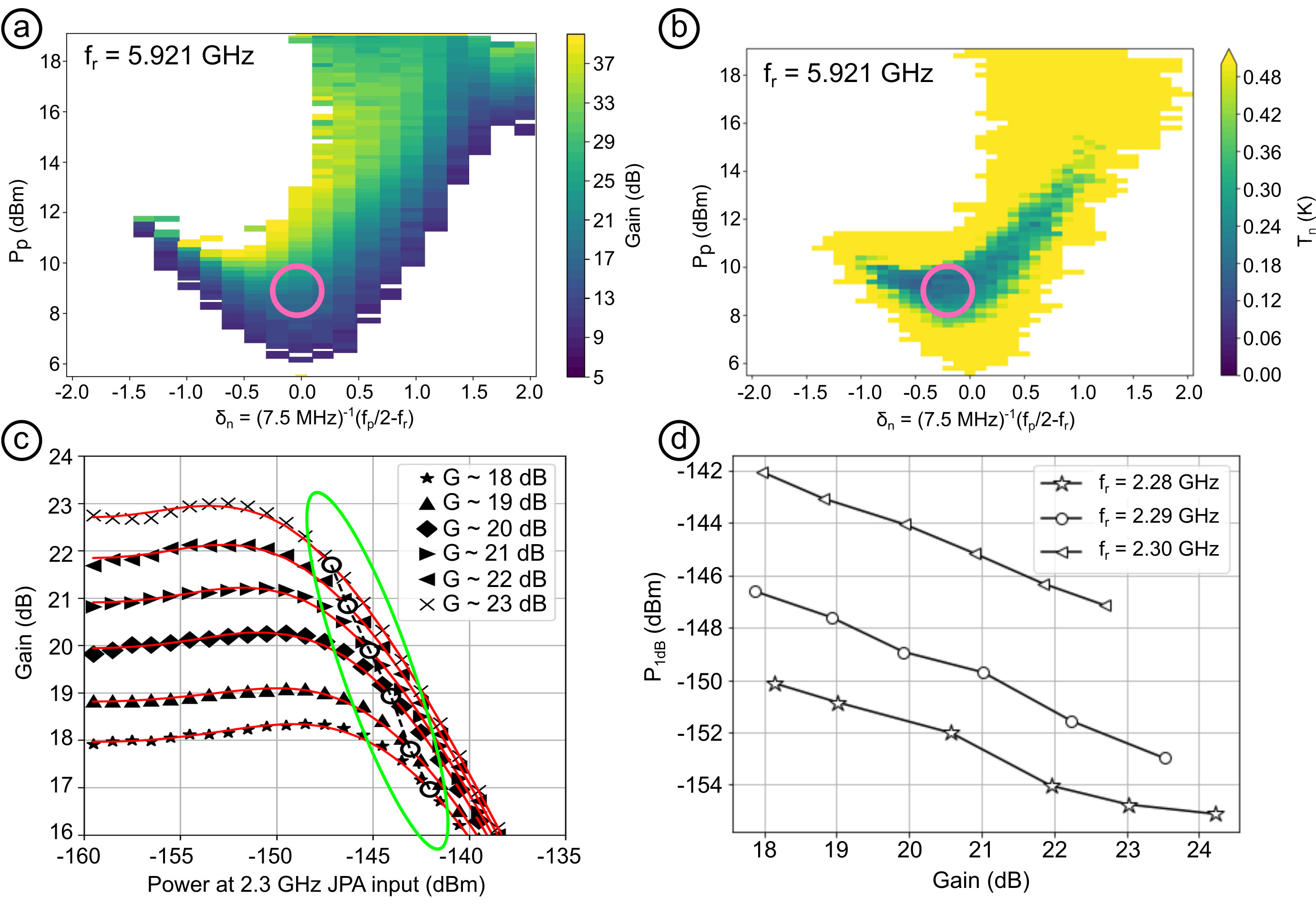}
\caption{\label{fig:Gain_and_Noise_paramap}Characterization of the JPAs. (a) and (b) Gain and noise parameter maps of the 6\,GHz JPA for $f_r=5.921$\,GHz. The area of optimal working points for $f_r=5.921$\,GHz is shown with pink circles. $\delta_n = 1/\text{BW}(f_p/2-f_r)$ is the normalized detunning, $\text{BW}$ refers to the JPA bandwidth. (c) Gain dependence of the 2.3\,GHz JPA on input power for one resonance frequency. (d) $P_{1\text{dB}}$ compression point for three resonance frequencies of the 2.3\,GHz JPA.}
\end{center}
\end{figure}

During paramap measurements, after each adjustment of \( i_b \), the resonance frequency \( f_r \) is established through phase measurements and subsequent parameter fitting. With the detuning defined as \( \delta = f_p/2 - f_r \), iso-gain contours exhibit a minimum in required pump power near \( \delta = 0 \), as depicted in Fig.~\ref{fig:Gain_and_Noise_paramap}(a). Initially, this minimum is observed at specific frequencies; however, it gradually shifts towards lower detunings, a phenomenon attributed to pump-induced shifts in the resonance  frequency~\cite{Krantz_2013}. As illustrated in Fig.~\ref{fig:Gain_and_Noise_paramap}(a), amplification can reach up to 40 dB.

Based on the data acquired from the gain paramap, we outlined an area of interest and measured a noise paramap [Fig.~\ref{fig:Gain_and_Noise_paramap}(b)] within that region. The displayed gain and noise paramap are for a 6 GHz JPA. As evident, we recorded an added noise level of around 150\,mK for this amplifier, slightly exceeding the quantum limit for a phase insensitive amplifier at this frequency $(\hbar\omega)/(2k_B)\approx$141\,mK~\cite{Haus62}.

For each JPA, we conducted numerous measurements at different  frequencies of interest to pinpoint the optimal configuration. The results presented in Figs.~\ref{fig:Gain_and_Noise_paramap}(c) and (d) showcase the outcomes of noise measurements for the 6\,GHz JPA using different settings at two distinct frequencies. The red vertical lines correspond to the quantum limits for each of these frequencies. As illustrated, for each frequency, we achieved parameters that align with very low noise levels, approaching the quantum limit.

\newpage

\subsection{P\textsubscript{1dB} measurement}\label{sec:P1dB}

Due to the nonlinear nature of the JPA, its gain will saturate beyond certain signal and pump amplitudes.
 A common method for assessing amplifier saturation involves determining the output power at which the gain decreases by 1 dB. This process involves driving the amplifier with a continuous tone at the desired signal frequency and progressively raising the input power level while tracking the output power.   $P_{1\text{dB}}$ is an indicator that the system deviates from the desired small-signal behavior.

Our measurement of $P_{1\text{dB}}$ was conducted at $\delta f = $ 1\,kHz across the entire JPA tunable range and various pump powers, each corresponding to different gains. The results are illustrated in Figs.~\ref{fig:Gain_and_Noise_paramap}(e) and (f). The power due to the axion signal, which is expected to be around $-180\,\text{dBm}$,  is significantly below the saturation power of the JPA. However, for a few  JPAs it was observed that thermal noise at the input can also induce saturation, altering the device behavior. As an example, when we considered a device with operating frequencies around $2.3\,\text{GHz}$ at gains exceeding $23 \, \text{dB}$, we observed saturation due to thermal noise when the NS temperature exceeded $120 \, \text{mK}$. While this may not render the device unusable at these frequencies, we need to  to careful during direct noise temperature measurements using a NS in these frequency and gain ranges~\cite{article:Kutlu21}.  

\subsection{Idler mode and added noise}
\begin{figure}[ht]
    \centering
    \includegraphics[width=0.9\textwidth]{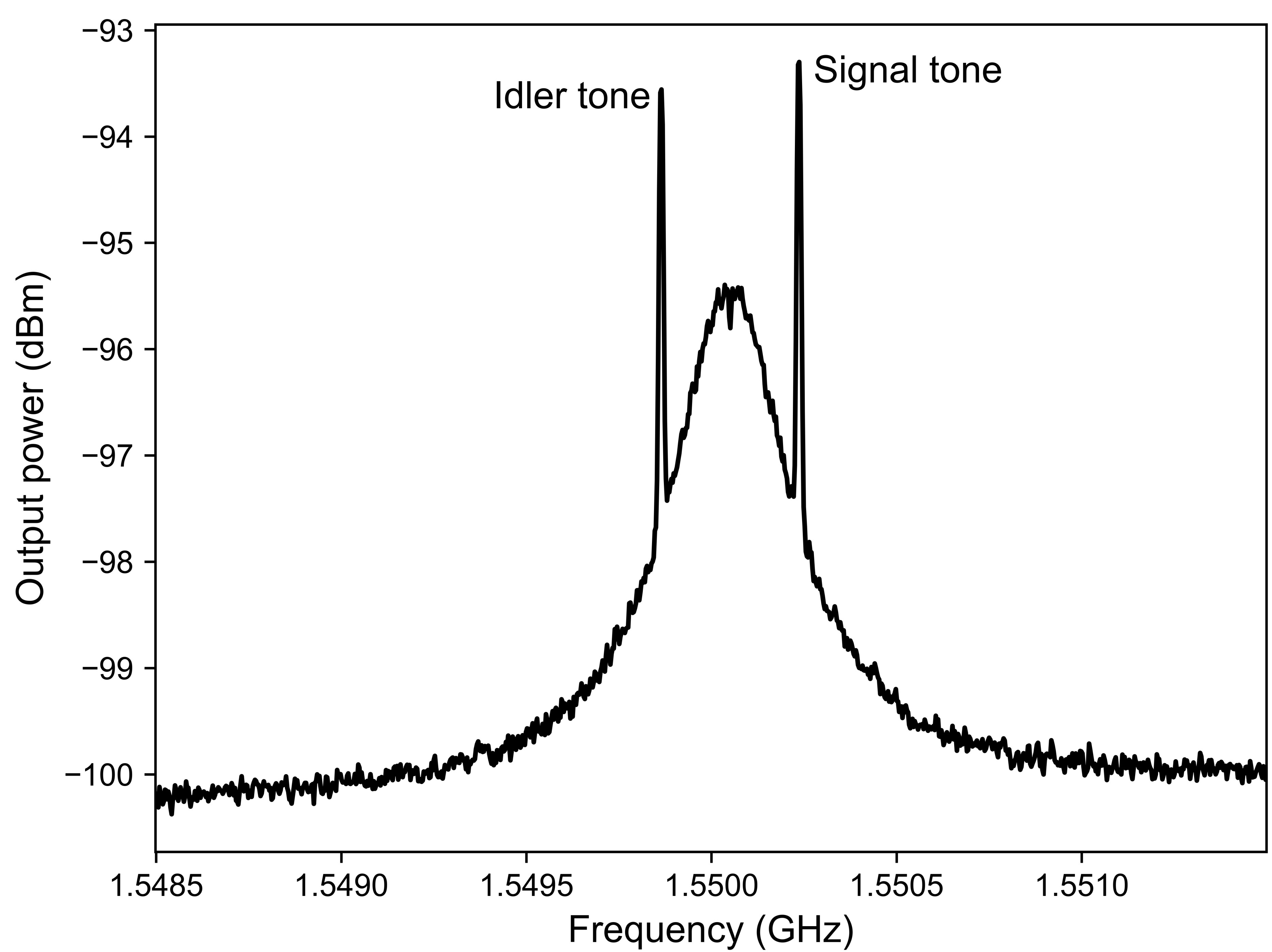}
    \caption{JPA output power measured by the spectrum analyzer including the amplified signal and its idler. A single signal tone is applied to the input of the JPA.}
    \label{fig:Idler}
\end{figure}

In the three-wave mixing non-degenerative mode, an idler tone appears with frequency $f_i =f_p-f_s$~(see Fig.~\ref{fig:Idler}). 
The amplification of the signal at the idler frequency is equal to 

\begin{equation}
\label{G_J}
    G_I = G_J - 1.
\end{equation}
 When the input signal $P_\text{IN}$ of the $f_s$ frequency is applied to the JPA, the output of the amplifier will have two power terms, corresponding to  $f_s$ and $f_i$ frequencies:

\begin{equation}
    P_\text{OUT} = G_J \left( P_\text{IN} + \frac{hf_s}{2} \right) + G_I \left( P_\text{IN} +\frac{hf_i}{2}\right).
\end{equation}
At $G_J \gg 1$, the added noise equals half a photon~\cite{Kuzmin83,Babourina-Brooks_2008}.

\subsection{Noise measurement}
To ascertain the noise temperature of the JPA, we utilize two steps. The first is based on the Y-factor method~\cite{Engen1970}, while the second involves measuring the noise spectrum both with the JPA on and off, a technique known as the spectrum comparison method~\cite{article:Friis1944}.

\subsubsection{Modified Y-factor}

To measure the noise temperature, we utilized a methodology akin to the well-established Y-factor method~\cite{Engen1970}, employing the NS discussed earlier in Section~\ref{sec:NoiseSource}. The Y-factor technique is commonly applied in the field of RF and microwave measurements, particularly for evaluating the noise figure of amplifiers or receivers. It involves calculating the ratio of output noise power observed when a system is connected to a calibrated hot load (high temperature)  $P^{\text{HOT}}$ to that observed when connected to a calibrated cold load (low temperature)  $P^{\text{COLD}}$. Mathematically, it is expressed as:

\begin{equation}\label{eq:Yfactor}
Y=\frac{P^{\text{HOT}}}{P^{\text{COLD}}}=\frac{ T_{\rm sys}^{\rm HOT}}{ T_{\rm sys}^{\rm COLD}}
\end{equation}

\noindent
The right part of Eq.~\ref{eq:Yfactor} reflects the fact that the noise temperatures $T_{\text{sys}}^{\text{HOT}}$ and $T_{\text{sys}}^{\text{COLD}}$ are proportional to the corresponding noise powers. The noise temperature of a system in both cases contains the amplifier-added noise $T_{\text{add}}$:

\begin{equation}\label{eq:Tsys}
T_{\text{sys}}^{\rm HOT} = T_n^{\rm HOT} + T_{\text{add}}, \quad T_{\text{sys}}^{\rm COLD} = T_n^{\rm COLD} + T_{\text{add}} 
\end{equation}

From Eqs.~\ref{eq:Yfactor} and \ref{eq:Tsys}, knowing $Y$, we obtain:

\begin{equation}
T_{\text{add}} = \frac{T_{n}^{\text{HOT}} - T_{n}^{\text{COLD}} Y}{Y - 1} ,\quad G_{\text{sys}} = \frac{P^{\rm HOT} - P^{\rm COLD}}{k_{B}BW(T_{n}^{\rm HOT} - T_{n}^{\rm COLD})}, \label{eq:t_add}
\end{equation}
where $BW$ is the bandwidth.
The advantage of this method is that it does not affect the precision of gain measurement $G_{\text{sys}}$.

\noindent
So, the original Y-factor method measures the noise temperature $T_{\text{sys}}$ at two points. To mitigate the influence of quantum effects on Johnson noise given by Eq.~\ref{eq:T_n}  and ensure control linearity, which can be compromised by JPA saturation, we have modified the standard method by measuring $T_{\text{sys}}$ at multiple temperature points.

For illustration we considered the measurement of noise with a 2.3\,GHz JPA. The noise power emerging from the system was measured with the SA with a 1 kHz resolution bandwidth. The power spectra were recorded at NS temperatures $T_s$ of 60, 120, and 180\,mK. The power values were converted into \ac{PSD} by dividing them by the resolution bandwidth of the SA. The pump power and resonance frequency were adjusted to achieve a JPA gain of approximately 20\,dB.

\begin{figure}[ht]
\begin{center}
\includegraphics[width=1\textwidth]{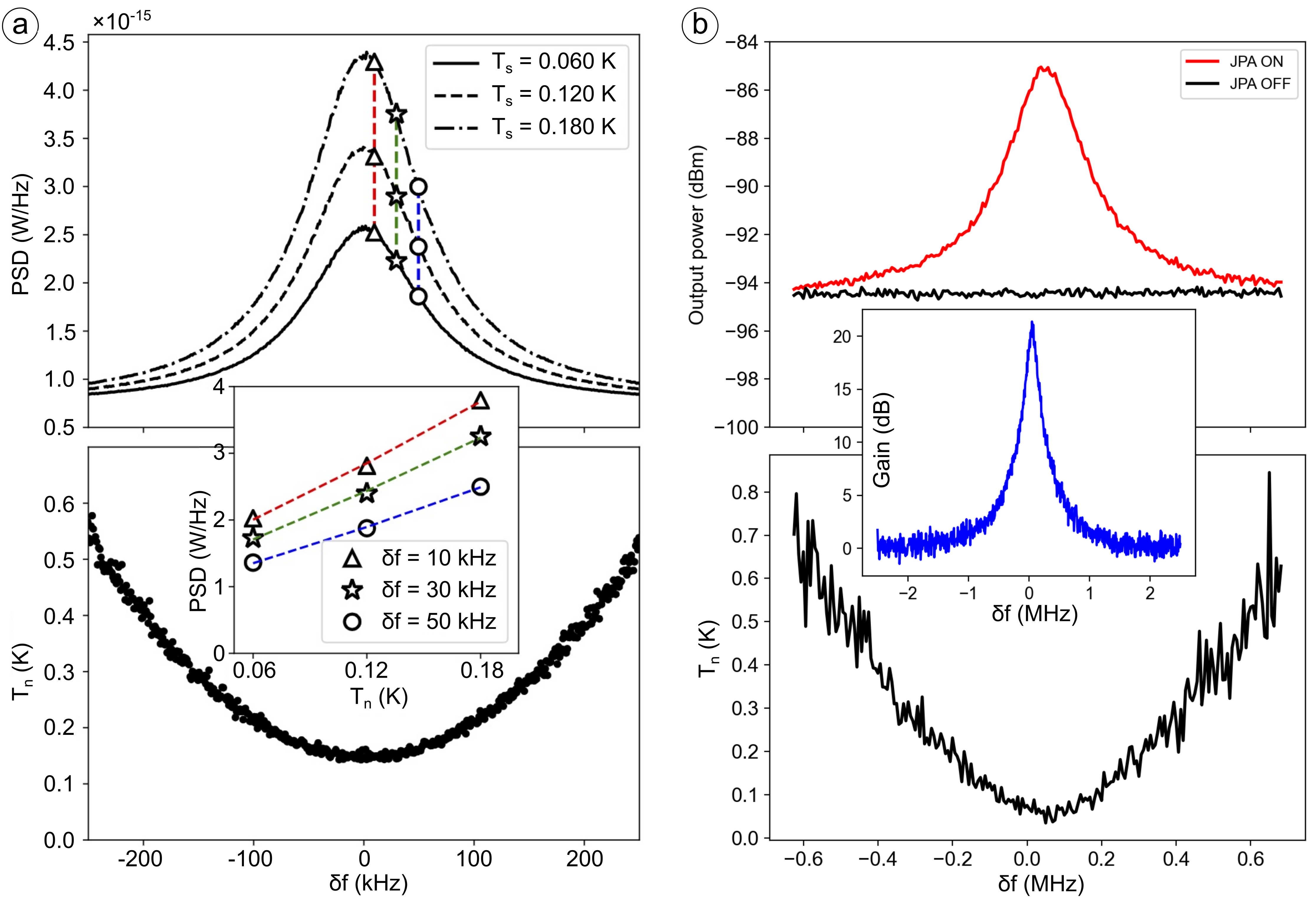}
\caption{\label{fig:Y-method_noise} Noise measurements from two different JPAs: (a) 2.3~\,GHz JPA noise measured using the Y-factor method. The upper plot shows the power spectral densities measured for $f_r = $ 2.305 GHz for three noise source temperatures at the optimal working point~\cite{article:Kutlu21}. The inset displays the PSD of noise for three frequencies used to calculate the noise temperature. The frequency dependence of the calculated noise temperature is shown in the bottom plot. $\delta f$ is a deviation from $f_r$, $\delta f = f - f_r$. (b) 1.58~\,GHz JPA noise measured using the spectrum comparison method. The upper plot shows the output power spectra for the JPA ON and JPA OFF. The inset displays the JPA gain curve. The noise temperature depending on the ofset frequency is shown in the bottom plot.}
\end{center}
\end{figure}

The lower limit on noise temperature for linear, phase-insensitive amplifiers is described by \cite{Clerk2010}:
\begin{equation}
T_\text{add}^\text{min} = \lim_{{S_0 \to 0}} \frac{S_n(f, T)}{k_B} = \frac{hf}{2k_B},
\end{equation}
which results in approximately 55.2\,mK at 2.3\,GHz. 
From the measurement in Fig.~\ref{fig:Y-method_noise}(a) we obtained a total added noise temperature of $T_\text{add}\approx120$\,mK, at the cavity physical temperature of 50\,mK. This cavity temperature at 2.3\,GHz frequency, according to Eg.~\ref{eq:T_n}, corresponds to a noise temperature of $\approx$~70\,mK. Therefore, the total system noise temperature is estimated to be $T_\text{sys} \approx 190$\,mK.
The lower bound for $T_\text{sys}$ is determined by the \ac{SQL}~\cite{Caves1982},
which is approximately 110\,mK at 2.3\,GHz. Therefore, for the 2.3\,GHz haloscope experiment, we obtain a total system noise of 1.7~SQL.  

\subsubsection{Spectrum comparison}\label{SpectrumComparison}

The second method we employ in the noise measurement is the spectrum comparison method. An advantage of this method is that it does not require the continuous usage of a NS, resulting in faster measurements as there is no need for changing the temperature. This technique is applicable because the JPA can be switched on and off, allowing noise measurements in both states. {The off state of the JPA corresponds to a condition where \( G_{\text{J}}^{\text{off}} \approx 1 \) and \( T_{\text{J}}^{\text{off}} \approx 0 \)} (absence of the pumping signal of the JPA and changing the bias current to move the JPA passive resonance away from the measured frequency range). The noise calculation consists of comparing the power spectra of the JPA in the ``on'' and ``off'' states and deducing \( G_{\text{sys}}^{\text{on}} \) and \( T_{\text{add}}^{\text{on}} \) from the results obtained with the JPA in the off state.

Initially, we characterized the HEMT noise by performing the Y-factor method with the JPA turned off. For our experiments around 1\,GHz the best noise we have measured with the RF-chain including HEMTs corresponds to 1\,K, which is approximately 40 times the half-quanta limit. \( G_{\text{sys}}^{\text{off}} \) and \( T_{\text{sys}}^{\text{off}} \) are measured using a VNA. After that the JPA is turned on, and \( G_{\text{sys}}^{\text{on}} \) is measured.   The JPA gain \( G_{\text{J}} \) is calculated using Eq.~\ref{eq:G_J}.
The VNA emits a signal of ample strength, rendering the noise negligible. By sweeping the frequency of the narrow-band signal, we can obtain the dependence of \(G_{\text{J}}(f_s)\).

The noise power spectrum on the output of AMP3 (see Fig.~\ref{fig:Fig2_main_diagram}), \(P_\text{sys}^{\text{on}}\), is recorded by a SA when 
the JPA is on.
Upon determining \(G_{\text{sys}}^{\text{on}}\), \(T_{\text{J}}\) can be estimated. Consequently, the total added noise of the JPA measured with the spectrum comparison method,  $T_{\rm add}^{\rm SC}$, is computed as:

\begin{equation}
    T_{\rm add}^{\rm SC} \approx \frac{P_\text{sys}^{\rm ON}}{P_\text{sys}^{\rm OFF}} \frac{T_\text{sys}^{\text{off}}}{G_{\rm J} } - \left( T_s + \frac{T_{\rm add}^{\rm off}}{G_{\rm J}} \right),
    \label{eq:total_added_noise}
\end{equation}
where $T_s$ is the stable physical temperature of the noise source while conducting the spectrum comparison method.

 As confirmed by our measurements, both methods yield results that are consistent within the measurement error. However, the spectrum comparison method allows achieving the same precision much faster. 
In recent experiments, we have only utilized the modified Y-factor method to measure the frequency dependence of the noise temperature of the HEMT amplifier chain. We have employed spectrum comparison methods to measure the noise of the JPA.

\subsection{Optimization of working point}

\subsubsection{Paramap Optimization}

Based on the JPA tests described in this section, we obtain an \ac{OWP} for each signal frequency $f_s$ and JPA gain of choice $G_J$ (usually close to 20\,dB by three parameters: $i_b$, $f_p$, and $P_p$. To determine the OWP, we used mainly two approaches. Historically the first method is based on the observation that the minimum noise at otherwise equal conditions coincides with the lowest $P_p$.

The JPA gain is a function of frequency, typically with a peak occurring at \(f_{s} = \frac{f_p}{2}\). Using $G_J$ to denote the peak gain value, the tuning is done by following the steps below:

\begin{enumerate}
    \item Tune \(i_b\) such that \(f_r\) is equal to the desired \(f_{s}\).
    \item Set \(f_p\) to \(2f_{s}\).
    \item Increase \(P_p\) until the desired gain is obtained.
    \item Repeat steps 2 and 3 for small deviations from \(f_r\) by changing \(i_b\).
    \item Pick a set of \(i_b\), \(f_p\), \(P_p\) that has the lowest \(P_p\) for the desired gain at \(f_{s}\).
\end{enumerate}

Before running the experiment, we conduct a series of preliminary measurements to determine the optimal values of the JPA parameters \((f_{s},G_J)\), which include $i_b$, $f_p$, and $P_p$.
Further details are described in~\cite{article:Kutlu23-LT29}.

We have also devised a method to heuristically find an optimized working point using the Nelder-Mead algorithm~\cite{article:NelderMead1965}.
This approach allowed us to identify a working point with a low noise temperature near a given initial seed~\cite{article:YounggeunKim2024}.

\subsection{CAPP JPA collection }

Since 2019, we have designed and tested over 40 JPAs for CAPP's axion search experiments. Parameters for some of these JPAs are presented in Table~\ref{tab:JPAs}.  

\begin{table}[h]
    \caption{Characteristics of some of the JPAs measured at physical temperature of 30\,mK.}
    \centering
    \begin{tabular}{|c|c|c|c|c|c|c|c|c|c|c|c|}
    \hline
      JPA   & \#1 & \#2 & \#3 & \#4 & \#5 & \#6 & \#7 & \#8 & \#9   \\ \hline
      Min Frequency, GHz     & 1.006 & 1.118  & 1.141 &1.2 &1.32&1.35 &1.4& 1.48 & 1.56    \\
      Max Frequency $f_\text{max}$, GHz    & 1.071 & 1.185 & 1.171 &1.251&1.401& 1.40 & 1.45&1.55 & 1.64   \\
     Tunable range, MHz      & 65 & 67 & 30 &51 &81&50&50 & 70 & 80   \\
     $T_\textrm{add}$, mK     & \textless90 & \textless90 & \textless60 &\textless95 &\textless80&\textless80 & \textless85 &\textless75 & \textless85 \\
     $T_n/(hf_\text{max}/(2k_B))$   & 3.5 & 3.2  &2.13 & 3.2 &2.38 &2.38& 2.44 & 2.01 & 2.15   \\
      Bandwidth  at 20dB  &  &  &  &  &  &  &  &  &    \\
             gain (kHz)  & \textgreater150 & \textgreater150& \textgreater100 &\textgreater200 & \textgreater150&\textgreater150 & \textgreater150 &\textgreater200  & \textgreater200  \\
    \hline
   \hline
      JPA   & \#10 & \#11 & \#12 & \#13 & \#14 & \#15 & \#16 & \#17 & \#18   \\ \hline
      Min Frequency, GHz     & 1.63 &  1.64 & 2.27 & 2.30  & 2.35&  2.35 & 3.7 & 5.1 & 5.5  \\
      Max Frequency $f_\text{max}$, GHz    & 1.73 &  1.74 & 2.31 & 2.45 & 2.5 &  2.5 & 3.85 & 5.4 & 6.0  \\
     Tunable range, MHz      & 100 &   100 &  40 & 15  & 150 &  150 & 120 & 300 &  500 \\
     $T_\textrm{add}$, mK     &   \textless80 & \textless80 & \textless120 & \textless130& \textless100 & \textless110 & \textless120 & \textless145 & \textless150 \\
     $T_n/(hf_\text{max}/(2k_B))$   & 1.93 & 1.92  & 2.16 & 2.32 & 1.67 & 1.83 & 1.3 & 1.12 & 1.04    \\
      Bandwidth  at 20dB  &  &  &  &  &  &  &  &  &    \\
             gain (kHz)  & \textgreater200 &  \textgreater150 & \textgreater100 & \textgreater200  & \textgreater200  & \textgreater100 & \textgreater300 & \textgreater1000 &  \textgreater1000 \\
    \hline
    \end{tabular}
    \label{tab:JPAs}
\end{table}

\section{Extension of amplification bandwidth}

Our primary challenge stemmed from the constrained tunable range imposed by the necessity to cool down our substantial 1.2-ton magnet using liquid helium. Consequently, we opted for a wet dilution refrigerator as a key component in our experimental setup to ensure safe superconducting magnet operations. Each cooldown of the dilution refrigerator consumes several hundred liters of liquid helium, and replacing the JPA requires several weeks of effort. Therefore, it is practical to match the amplifier's bandwidth with the cavity's, which is approximately 300 MHz. This can be achieved by employing a split-band technique, wherein each amplifier operates within its designated frequency band.

\subsection{Cold multiplexer}
\begin{figure}[ht]
\begin{center}
\includegraphics[width=1\textwidth]{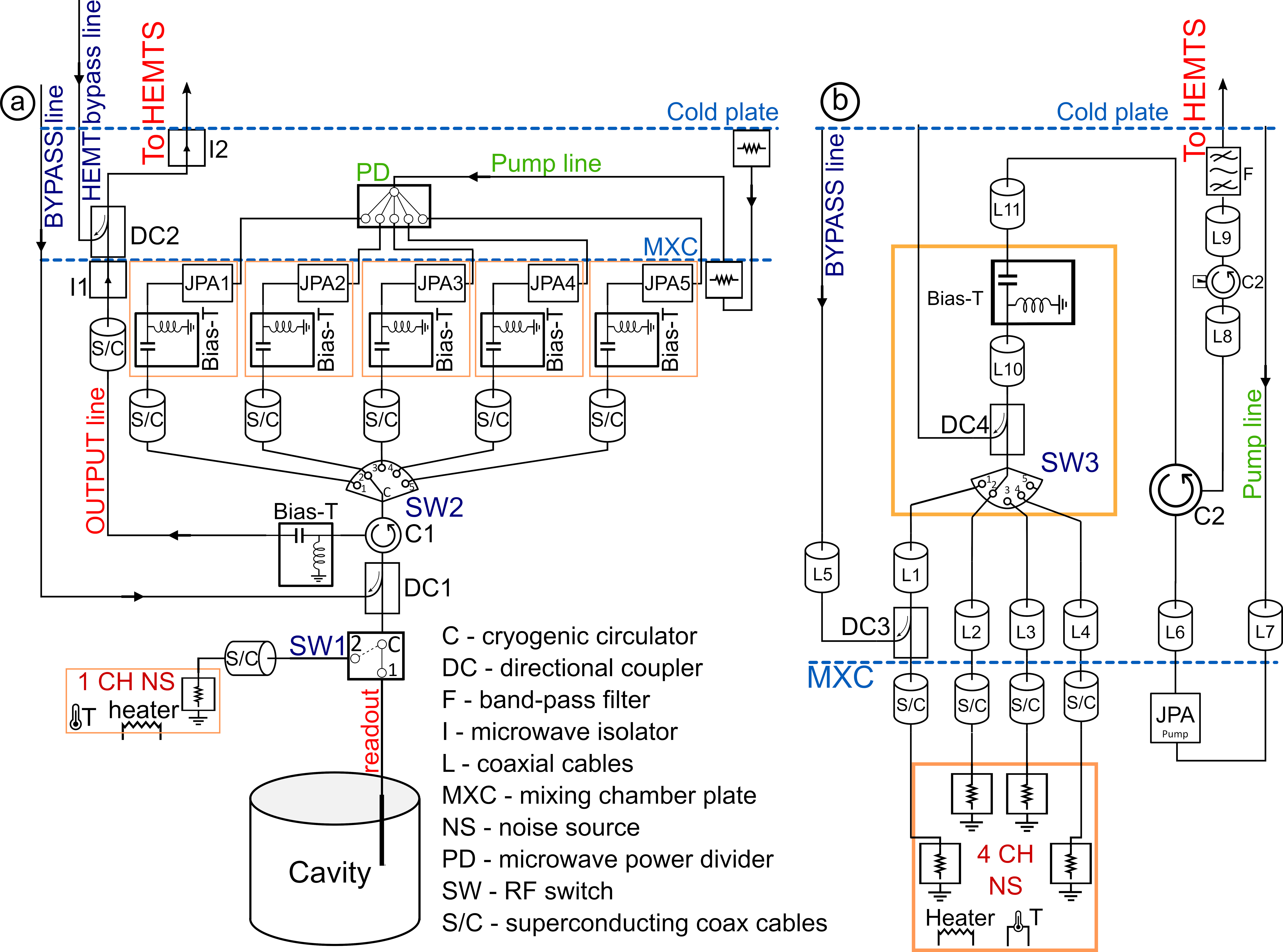}
\caption{\label{fig:MUX_diagram} JPA multiplexing setup at the MXC: (a) The diagram of the detection chain in the axion search experiment. The setup consists of the cavity readout line and five JPAs measuring line. The switching is provided by the RF switch SW1. Compared with the JPA readout~\ref{fig:Fig2_main_diagram}, a 5-position RF switch SW2 is added, which connects five JPAs through superconducting cables and protective bias-tees. The setup utilizes one channel noise source 1 CH NS. The pump signal is split using the power divider PD. The output is followed by an additional bias-tee, cryogenic microwave isolator I1, directional coupler DC2, and cryogenic isolator I2. (b) Setup for characterizing the transmission of each channel of the 5-position RF switch.}
\end{center}
\end{figure}

Initially we considered a measurement setup with a microwave RF switch as the simplest means for extending the scanning frequency range of the RF chain.
The multiplexing design of the multiple JPA readout scheme was based on the five-position RF switch, shown as \text{SW2} in Fig.~\ref{fig:switch_shield}a.
This way, we designed the setup, such that each channel of the RF switch is connected to the one of five JPAs, utilizing a common output, see Fig.~\ref{fig:MUX_diagram}a.
The axion signal detection chain starts from the cavity, connecting to the RF switch \text{SW1}, which is used to switch between the cavity (scanning for axions) and the noise source (noise measurements). 
The first directional coupler \text{DC1} connects to the \text{BYPASS line}, which is used for the characterizing of the passive resonances of each JPA.
Similar to the circuit diagram shown in Fig.~\ref{fig:Fig2_main_diagram} the cryogenic circulator \text{C1} is connected to the JPA.
The five-position RF switch \text{SW2} switches between JPAs according to the operating frequency range in axion search experiment.
The pump line for each JPA utilizes a power divider \text{PD}, and the pump frequency and power are set according to the active JPA. 
The output of multiplexed JPA circuit is connected to the microwave isolator \text{I} and an additional directional coupler \text{DC2}. 
\text{DC2} is needed to perform additional measurements of the HEMT chain, see \text{HEMT bypass line} in Fig.~\ref{fig:MUX_diagram}a, to measure the losses in the output chain of the JPA, and to precisely characterise the noise temperature.
In order to characterize the differences in transmission for each channel of the RF switch, we performed additional measurements with the circuit diagram shown in Fig.~\ref{fig:MUX_diagram}b.
We used a four channel noise source~\cite{article:Ivanov23-LT29}, shown as \text{4 CH NS}, with 4 identical cryogenic wideband terminators connected to the five-position RF switch.
The common port of the RF switch is connected to the cryogenic circulator and to the JPA.
We performed measurements of the JPA noise temperature for each of the channels and did not observe 
any differences.

In our early tests, we encountered an issue where a JPA was damaged. Further investigation revealed the potential for charge accumulation on the switch electrodes during mechanical movement. This accumulated charge could discharge at the moment of switching, generating current and voltage pulses that posed a threat to the amplifiers. At mK temperatures, leakage current can be exceptionally low due to charge carriers freezing in insulating materials, resulting in prolonged charge retention on the electrodes. Once we recognized this issue, we implemented bias tees, shown as \text{Bias-T} in Fig.~\ref{fig:MUX_diagram}a, to safely discharge the switch, enabling us to use this method for switching between several JPAs without causing damage.
Another challenge associated with this method is the necessity to supply a substantial current ($\approx$250\,mA) inside the fridge to facilitate channel switching and reset the RF switch. 
To achieve this, we would need to modify the fridge by adding more low-resistance cables. 
As of now, this modification has been deferred.
After addressing these issues, the proposed circuit scheme, which utilizes a 5-position multiplexer combined with the parallel-series connection design described in the next chapter, will allow for the expansion of the operating frequency range beyond 1 GHz.

\subsection{Series connection of JPAs}

\begin{figure}[ht]
\begin{center}
\includegraphics[width=1\textwidth]{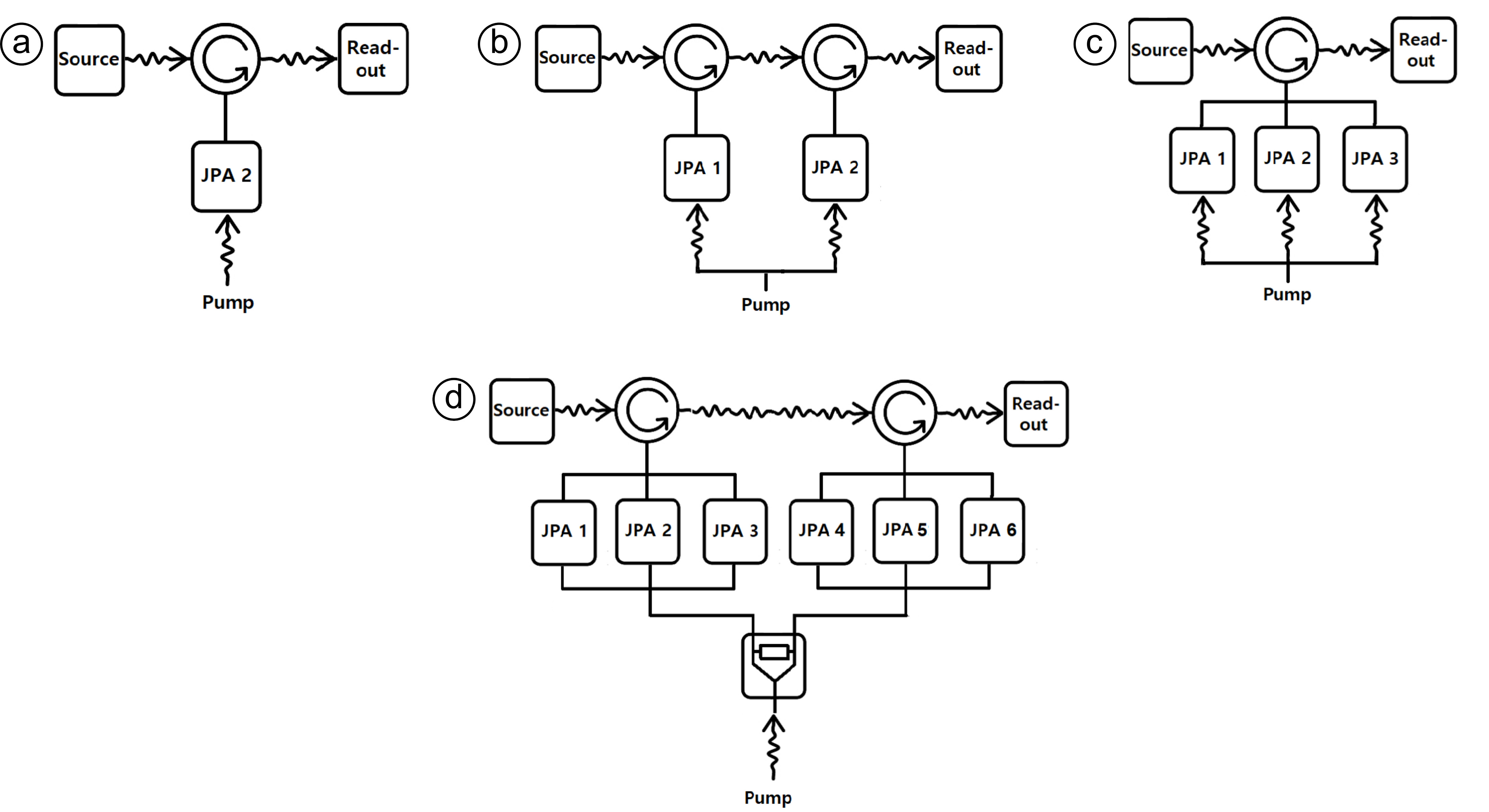}
\caption{\label{fig:JPA_all_connections} Simplified amplifier schematics. (a) single JPA connection, (b)  serial connection, (c) parallel and (d) parallel-serial connections.}
\end{center}
\end{figure}

In order to increase the bandwidth of our readout, we introduced a series connection involving an additional JPA alongside the first JPA [JPA1 and JPA2, see Fig.~\ref{fig:JPA_all_connections}(b)]. During measurements, only one JPA operates at a given time, while the other JPA remains detuned from its operational frequency using a DC flux bias. This configuration seamlessly integrates into the readout chain via the second circulator, resulting in additional losses of approximately 0.4\,dB at frequencies around 1\,GHz. These losses, corresponding to less than 10\% of the signal, lead to a minor acceptable increase in noise temperature. 

The series connection has enabled us to simultaneously use two JPAs within the same cooldown cycle, a setup we have employed in our test fridges since 2021.

\subsection{Parallel connection of JPAs}

\begin{figure}[ht]
    \centering
    \includegraphics[width=1\textwidth]{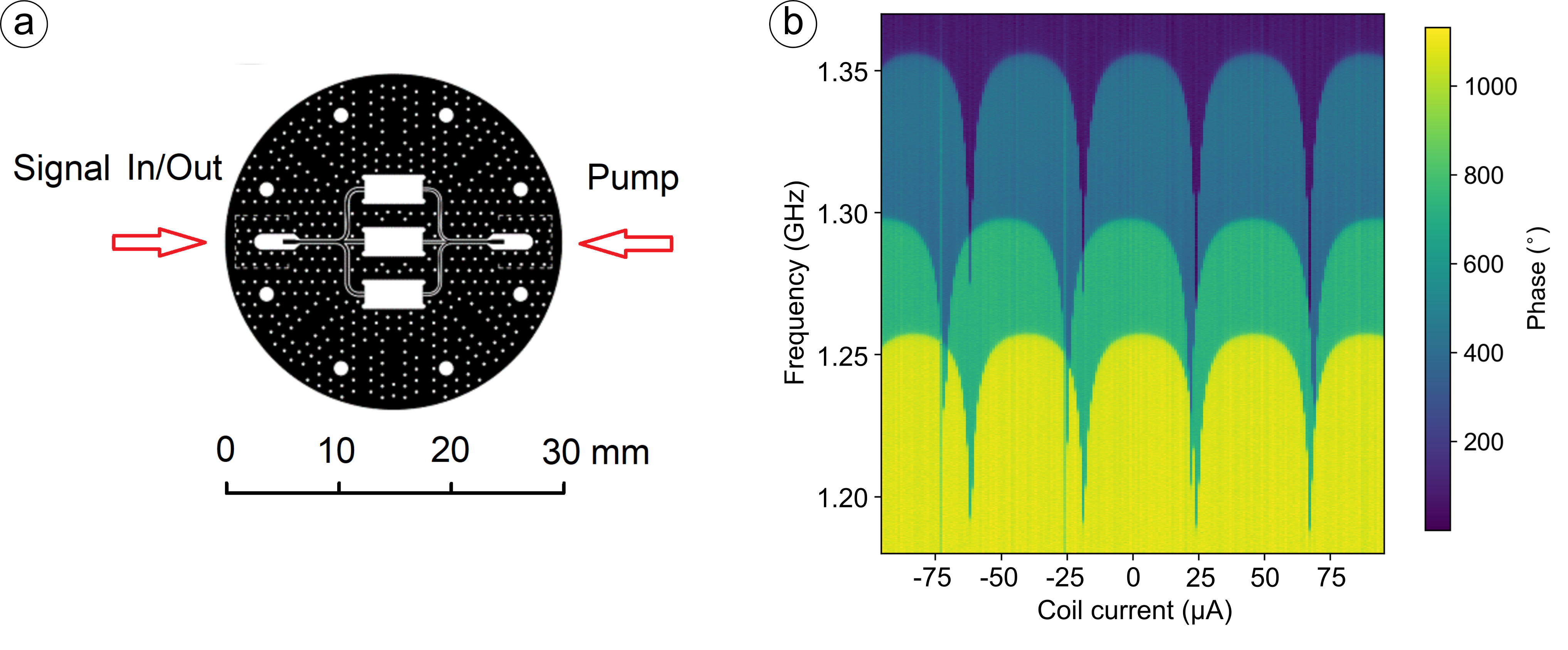}
    \caption{{\textcolor{black}{Parallel connection of three JPAs~\cite{article:Uchaikin-LTD20}. (a) PCB for the parallel connection of 3 JPAs. The DC flux bias is common for all of the JPAs and is created by a 200-turn coil wound with superconducting wire around the PCB holder (not shown). (b) Dependence of the resonance frequency on bias current for three JPAs connected in parallel, with the JPA pump switched off.}}}
    \label{fig:Gorynych}
\end{figure}

The second idea for the bandwidth extension is to connect multiple JPAs with different frequencies on a single PCB, which appeared feasible given that the wavelength of the signal at 1--2\,GHz significantly exceeds the length of the PCB traces of a few mm. The dimensions of the PCB we employ comfortably accommodate three chips, as depicted in Fig.~\ref{fig:Gorynych}(a). 

The initial measurements using this parallel configuration demonstrated that the JPAs can operate independently, exhibiting noise and gain characteristics similar to those in a single-JPA operation configuration~\cite{article:Uchaikin-LTD20}. However, certain challenges may arise because the two JPAs on opposite sides of the PCB [see Fig.~\ref{fig:Gorynych}(a)] share the same sensitivity to DC bias, causing their flux-to-frequency characteristics to change similarly. In Fig.~\ref{fig:Gorynych}(a), we can see three JPAs mounted on a PCB in order of their \ac{HPRF}: 1.28, 1.295, and 1.36 GHz. In Fig.~\ref{fig:Gorynych}(b), one can see the dependence of the resonance frequency on  bias current for the three JPAs connected in parallel, with the JPA pump power switched off. As we can observe, there is similarity in the DC flux bias sensitivity of the JPAs with maximum resonance frequencies of 1.28 and 1.36 GHz. This similarity can potentially lead to interference in the most sensitive low-frequency range, a matter which is discussed further in Sec.~\ref{Leo II}.

\subsection{Parallel-serial connection}  

Combining the two ideas and developing a parallel-serial JPA readout, where two parallel configurations are linked in series was the next logical step~\cite{article:Kim-LTD20}. This configuration was initially evaluated in an RF assembly named~``Leo-I'', which contained all RF components intended to be placed at MXC. The goal was to test the entire assembly in a dry fridge first, before moving it to the wet CAPP-MAX fridge without any modifications.

Before the dry fridge test, we arranged three JPAs in parallel with HPRFs of 1.07, 1.12, and 1.13\,GHz in the first holder based on the Onion shield, while the second holder accommodated a single JPA of 1.185 GHz. Unfortunately, the middle JPA of 1.12\,GHz, which was initially mounted and characterized on a single-JPA PCB version, was damaged during the transfer to the three-JPA PCB setup. Subsequent tests on the remaining three JPAs confirmed that all of them exhibited low noise levels while maintaining a required gain of 20\,dB. 

\begin{figure}[ht]
\begin{center}
\includegraphics[width=1\textwidth]{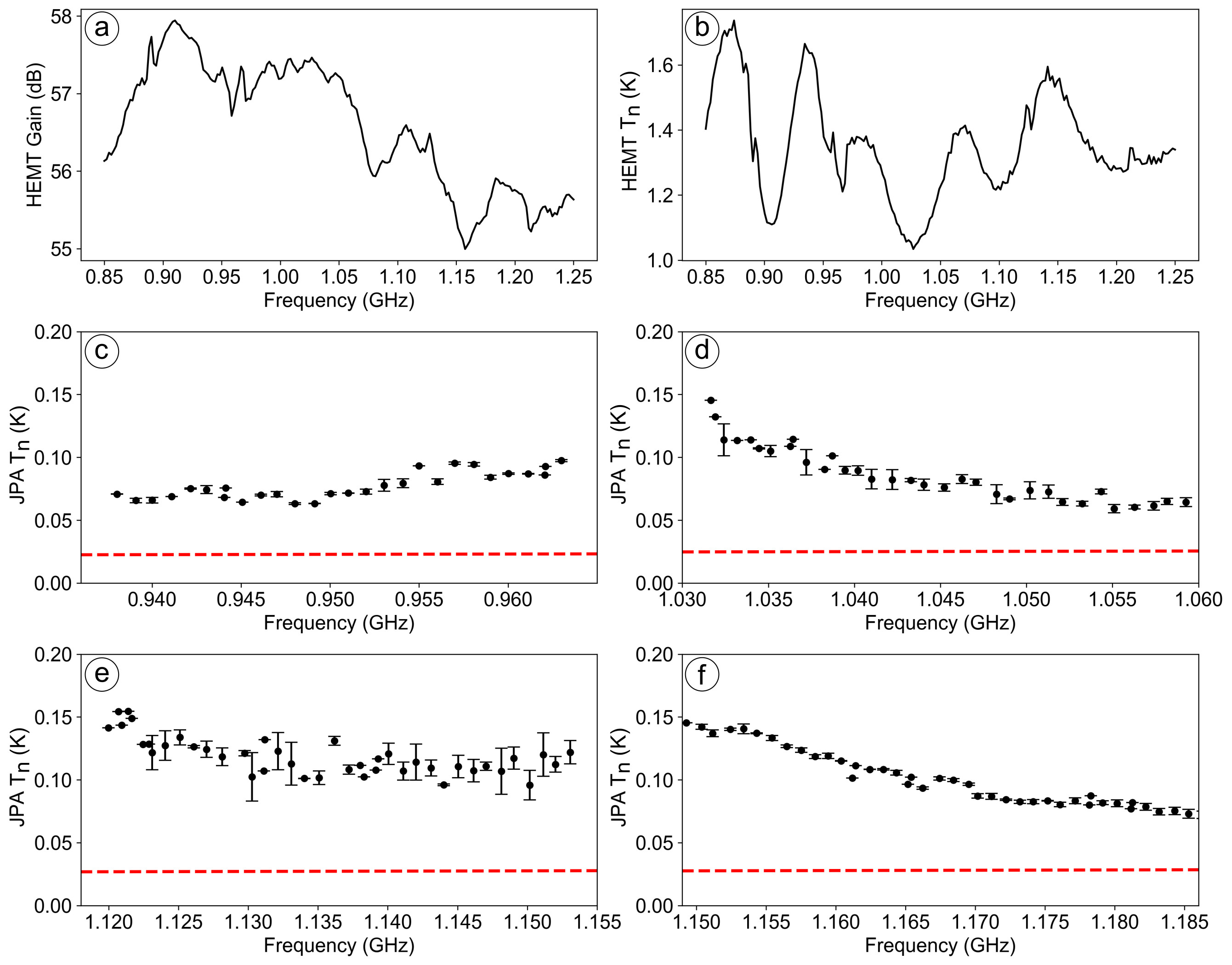}
\caption{\label{fig:Leo1} Characteristics of multi-JPA amplifier based on Leo~I assembly. (a) HEMT amplifier gain, (b) HEMT amplifier noise. (a) and (b) were measured with all JPAs OFF. (c)--(f)~Noise temperatures of JPAs 1--4, respectively. The red dashed line corresponds to the quantum limit for the amplifier added noise.}
\end{center}
\end{figure}

The complexity introduced into the readout chain led to some uneven frequency response and variations in HEMT noise temperature measured with the JPA off, ranging from 1.09 to 1.7\,K [see Fig.~\ref{fig:Leo1}(a) and (b)]. This cannot be entirely eliminated due to inherent phase delays in cables and the theoretical impossibility of achieving a perfect match for three-port non-reciprocal components within the chain. However, we accepted this limitation because at a JPA gain of 20\,dB, the effect of the next HEMT amplification stage on noise is reduced by a factor of 100, resulting in only a small impact on the system noise temperature.

In Run~6 of the CAPP-MAX experiment, as mentioned earlier, three JPAs were in operation. Two of them were situated within one holder, while the third was mounted in a separate holder. The DC flux bias was shared by connecting the DC flux coils of both holders in series. 
Further details of the measurements conducted during Run~6 of the CAPP-MAX experiment utilizing this amplifier can be found in~\cite{article:PRX2024, article:Kim23-LT29}.

\subsection{JPA interference in split band amplifier}\label{Leo II}

For Run~7 of the CAPP-MAX experiment, we prepared a new "Leo II" assembly, comprising a full set of 6 JPAs with HPRFs of 1.25, 1.30, 1.36, 1.40, 1.45, and 1.50 GHz in a parallel-serial connection as shown in Fig.~\ref{fig:JPA_all_connections}(d).  To minimize possible interference, we ordered the JPAs to avoid close operating frequencies among neighbors. In the first holder, we positioned JPAs starting with the lowest HPRF (1.25 GHz), followed by 1.36 GHz, and finally 1.45 GHz. The second holder followed a sequencing of 1.30 GHz, 1.40 GHz, and 1.50 GHz.

\begin{figure}[ht]
\begin{center}
\includegraphics[width=0.9\textwidth]{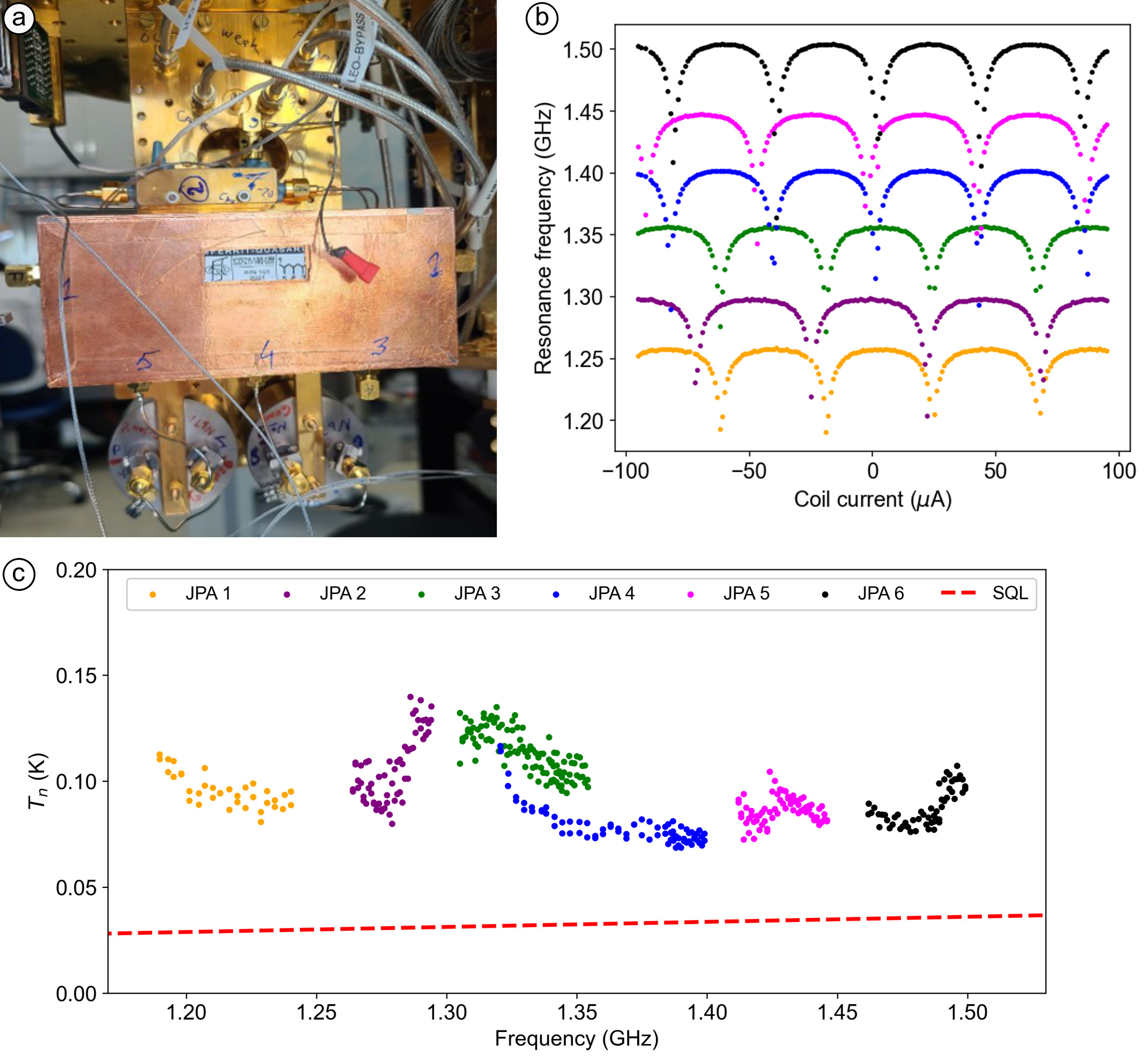}
\caption{\label{fig:Leo2} Six-JPA assembly (Leo II). (a) Assembly installed at the mixing chamber plate of the dry dilution refrigerator. (b) Flux-sweep curves of the six JPAs. (c) Added noise temperature depending on the frequency for the six JPAs. The red dashed line corresponds to the quantum limit for the amplifier added noise.}
\end{center}
\end{figure}

Tests of this assembly (Fig.~\ref{fig:Leo2}) in the dry fridge revealed characteristics similar to those of Leo~I. These JPAs allowed us to cover a frequency range of about 300\,MHz, spanning from 1.2 to 1.5\,GHz. However, interference between some of the JPAs at their lowest operating frequencies emerged during testing.

This setup  was tested with a series connection of DC bias lines for the two JPA holders. This configuration posed challenges, particularly in achieving detuning of the JPA resonance frequencies from each other, especially within the narrow range of the lowest operating frequencies where the sensitivity of JPAs to interference is most pronounced.
Due to the use of dc bias coils with the same design and almost identical JPA flux sensitivities, leveraging the $\Phi_0$ periodicity of SQUID characteristics was not feasible. Utilizing another branch of the flux-to-frequency characteristic would have resulted in the same interference problem. 
We decided not to modify the design due to the risk of damage.

While it would have been feasible to utilize different twisted pairs to supply DC flux bias current for the second holder, we deliberated several ideas to tackle this issue, with the objective of minimizing the interval between runs without requiring time-consuming fridge modifications.

Another approach involved connecting the DC-flux coils in parallel, with small resistors of different values in series with the coils in each branch. In this configuration, the coils will have different sensitivity to input current, but there is still a probability of interference for JPAs in different holders due to overlap at lower frequencies.

Furthermore, we developed and implemented a new circuit design using Schottky diodes. This circuit allowed us to independently apply two DC flux biases to two JPA holders using the same twisted pair of wires. It achieved this by separating the flux biases using currents of different directions to bias one or the other JPA holder. The circuit featured a diode rectifier operating at the 4\,K stage of the fridge to address heat dissipation in the diodes~\cite{article:Ko2023}.

A potential issue arose during the operation of GaAs diodes in proximity to a strong magnetic field due to the field's effect on the diode's performance. 
This effect could lead to changes in properties depending on the orientation of the diode in the magnetic field~\cite{article:Sun2004}. 
To mitigate this concern, we replaced each of the two diodes with three diodes, each oriented along different orthogonal Cartesian axes~\cite{article:Ko2023}.

After Run~6 of our axion experiment, we re-accessed the fridge design. 
We decided to upgrade our DC lines and implement a twisted pair to control the second JPA holde independently.
As a result, for Run~7, we were able to forego the use of this diode circuit, postponing its installation.

Before Run~7, we conducted a 6-JPA assembly (Leo-II) measurements and confirmed their noise characteristics [see Fig.~\ref{fig:Leo2}(c)] in the experiment fridge.


\section{Conclusions}
In this paper, we have presented an overview of the characterization and implementation of JPAs at CAPP, detailing the processes involved in achieving high sensitivity in axion detection.
Drawing from the established flux-driven JPA technology developed by RIKEN and the University of Tokyo, we successfully adapted these amplifiers for the CAPP haloscope search experiments through several key developments:

\begin{enumerate}
    \item We investigated and devised methods to operate flux-driven JPAs across a continuous frequency range. This involved introducing the paramap method to determine the operating points aimed at achieving specific gains and minimal noise. We  established a test setup based on a dilution fridge, complemented by a computer-controlled system for JPA testing and experiment execution.

    \item We introduced two methods for extending the effective amplifier frequency range, utilizing series and parallel connections. Our experiments with multiple JPA configurations, including the 6-JPA assembly, have demonstrated the efficacy of our approaches in achieving a broad frequency range. By selecting JPA sequences and employing different techniques, we successfully minimized interference and maximized sensitivity without compromising noise characteristics. By connecting 6 JPAs, each with an operating frequency band of 48--52\,MHz and different central frequencies, we expanded our coverage to approximately 300\,MHz, which is almost six times the range of a single JPA. It enhances our scanning capabilities by minimizing downtime and helium loss during cooldowns for the CAPP-MAX system~\cite{article:PRX2024}.

    \item We introduced a nested shield based on superconductors and Cryophy, enabling operation in high magnetic fields up to 0.1\,T without saturating the Cryophy and achieving a shielding factor of $10^5$--$10^6$ with residual fields below 100\,nT for stable JPA operation. A patent application for this innovation is currently underway.

    \item Our JPA development has facilitated the implementation of JPA-based amplifiers in various CAPP experiments, as described in~\cite{article:CAPP-PACE-JPA,12TB-PRL, Sagitarius,article:Kutlu23-PATRAS, article:Younggeun-Kim2023, article:PRX2024}, and has allowed the CAPP to contribute to the global axion search effort.
    
\end{enumerate}
These developments pave the way for further breakthroughs in future CAPP experiments.
\section{List of Acronyms}

\begin{acronym}[MPCCCCC] 
\acro{ADMX}{Axion Dark Matter Experiment}
\acro{BF}{BlueFors}
\acro{CAPP}{Center for Axion and Precision Physics Research}
\acro{CP}{cold plate}
\acro{CPW}{coplanar waveguide}
\acro{DAQ}{data acquisition}
\acro{DR}{dilution refrigerator}
\acro{DFSZ}{Dine-Fischler-Srednicki-Zhitnitskii}
\acro{HAYSTAC}{Haloscope At Yale Sensitive To Axion Cold dark matter}
\acro{HEMT}{high-electron-mobility transistor}
\acro{HPRF}{highest passive resonance frequency}
\acro{IBS}{Institute for Basic Science}
\acro{JPAs}{Josephson Parametric Amplifiers}
\acro{KAIST}{Korea Advanced Institute of Science and Technology}
\acro{KSVZ}{Kim-Shifman-Vainshtein-Zakharov}
\acro{LNA}{low noise amplifier}
\acro{MXC}{mixing chamber plate}
\acro{NS}{noise source}
\acro{OS}{Onion Shield}
\acro{OWP}{optimal working point}
\acro{PCB}{printed pircuit board}
\acro{PID}{proportional-integral-derivative}
\acro{PSD}{power spectral densities}
\acro{RF}{radio-frequency}
\acro{RQC}{RIKEN Center for Quantum Computing}
\acro{RT}{room temperature}
\acro{SA}{spectrum analyzer}
\acro{SM}{Standard Model}
\acro{SQL}{standart quantum limit}
\acro{SQUID}{superconducting quantum interference device}
\acro{VNA}{vector network analyzer}
\acro{VSWR}{voltage standing wave ratio}

\end{acronym}
\section*{Conflict of Interest Statement}

The authors declare that the research was conducted in the absence of any commercial or financial relationships that could be construed as a potential conflict of interest.

\section*{Author Contributions}

SVU led the development of the amplifier and proposed the idea of parallel connection of the JPAs. JMK proposed the idea of series connection of the JPAs. AFL and YN developed and produced the JPAs. SVU, CK, JMK, and BII developed the test setup. SO and SVU contributed to sample preparation. CK and JMK prepared the software control. CK, JMK, and BII performed the tests in the test fridge. JMK, BII, and SA tested the setup in the axion experiment fridge. BII, MK, and SVU contributed to component tests. BII, MK, SVU, JMK, and SA prepared the figures. SVU, CK, JSK, JMK, BII, AFL, YN, SA, and YKS contributed to the interpretation of the results. YKS led the axion search in CAPP. SVU wrote the manuscript, and CK, JSK, JMK, BII, AFL, YN, MSK, SA, and YKS contributed to text refinement. All authors agree to be accountable for the content of the work.


\section*{Acknowledgments}
This work was supported by the Institute for Basic Science (IBS-R017-D1) and  
JSPS KAKENHI (Grant No.~JP22H04937). Arjan F. van Loo was supported by a JSPS postdoctoral fellowship. The authors thank Dr. Seongtae Park for his assistance with the PCB development.

\bibliographystyle{unsrt}

\bibliography{main}

\end{document}